\pdfoutput=1

\documentclass[structabstract]{aa}  
\usepackage{graphicx}
\usepackage{natbib}
\usepackage{xspace}

\usepackage{txfonts}
\begin{document}

   \title{A new nearby pulsar wind nebula overlapping the RX J0852.0$-$4622 supernova remnant}

 \titlerunning{A new nearby PWN overlapping the RX J0852.0$-$4622 SNR}

\authorrunning{Acero et al.}

   \subtitle{ }

   \author{
   F. \,Acero\inst{1,2}, 
Y. \, Gallant\inst{1},
J. \, Ballet\inst{3},
M. \, Renaud\inst{1}
\and R. \, Terrier\inst{4}
          }

   \institute{
   LUPM, CNRS/Universit\'e Montpellier 2
\email{facero@in2p3.fr}
\and
NASA Goddard Space Flight Center, Greenbelt, MD 20771, USA
\and
AIM, CEA/CNRS/Universit\'e Paris Diderot-Paris 7 
\and 
APC, CNRS/Universit\'e Paris Diderot-Paris 7       
             }

   \date{Received 2012 November 27; accepted 2012 December 17}

% \abstract{}{}{}{}{} 
% 5 {} token are mandatory
 
  \abstract
  % context heading (optional)
  % {} leave it empty if necessary  
   {   Energetic pulsars can be embedded in a  nebula of relativistic leptons which is powered by the dissipation of the rotational energy of the pulsar.   
   The object PSR\,J0855$-$4644 is an energetic and fast-spinning pulsar ($\dot{E} = 1.1 \times 10^{36}$ erg/s, $P$=65 ms) discovered near the South-East rim of the supernova remnant (SNR) RX\,J0852.0$-$4622 (aka Vela Jr) by the Parkes multibeam survey. The position of the pulsar is in spatial coincidence with an enhancement in X-rays and TeV $\gamma$-rays, which could be due to its putative pulsar wind nebula (PWN).  }
  % aims heading (mandatory)
   {The purpose of this study is to search for diffuse non-thermal X-ray emission around PSR\,J0855$-$4644 to test for the presence of a PWN and to estimate the distance to the pulsar.}
  % methods heading (mandatory)
   { An X-ray  observation was carried out with the \textit{XMM-Newton} satellite to constrain the properties of the pulsar and its nebula.   The absorption column density derived in X-rays from the pulsar  and from different regions of the rim of the SNR was compared with the absorption derived from the atomic (HI) and molecular ($^{12}$CO) gas distribution along the corresponding lines of sight to estimate the distance to the pulsar and to the SNR. }
  % results heading (mandatory)
   {The observation has revealed the X-ray counterpart of the pulsar together with surrounding extended emission thus confirming the existence of a PWN. 
   The comparison of column densities provided an upper limit to the distance of the pulsar PSR\,J0855$-$4644  and the SNR RX J0852.0$-$4622  (d$\leq$ 900 pc).  Although both objects are at compatible distances, we rule out that the pulsar and the SNR are associated. With this revised distance,  PSR\,J0855$-$4644 is the second most energetic pulsar, after the Vela pulsar, within a radius of 1 kpc and could therefore contribute to the local cosmic-ray e$^{-}$/e$^{+}$ spectrum.
 }
  % conclusions heading (optional), leave it empty if necessary 
   {}

   \keywords{ Pulsars: general; ISM: supernova remnants; ISM: individual objects: (PSR\,J0855$-$4644, RX J0852.0$-$4622);
               }

   \maketitle
%
%________________________________________________________________

\section{Introduction}

The Vela region (approximately centered on $\ell = 265\degr$, $b=-1\degr$) is the host of several remnants of stellar explosions
including the eponymous supernova remnant (SNR). The Vela remnant  is among the brightest and most extended X-ray sources in the sky
and it is also the closest  SNR currently detected \citep[the distance to the remnant's pulsar has been measured to be 290 pc;][]{dodson03}.
 With its large angular size (8$\degr$ in diameter) the object is overlapping several remnants such as the Puppis A SNR and  RX J0852.0$-$4622. 
In the X-ray spectral domain, this complex region is dominated by the Vela and Puppis remnants in soft X-rays  
and the RX J0852.0$-$4622 SNR (aka Vela Jr)  becomes apparent only above 1 keV \citep{aschenbach98}.

The Parkes multibeam radio survey discovered a pulsar, PSR\,J0855$-$4644,
lying on the South-East rim of the RX J0852.0$-$4622 SNR  \citep{kramer03}. 
This young and energetic pulsar (characteristic age $\tau_{\rm c}$=140 kyrs, $\dot{E} = 1.1 \times 10^{36}$ erg/s and P=65 ms)
  is in spatial coincidence with an enhancement in the X- and $\gamma$-ray maps
 \citep[presented in][respectively]{aschenbach99,ah07-VelaJr}, which could be the signature of its pulsar wind nebula. 
Based on the radio dispersion measure of the pulsar  \citep[DM=239 cm$^{-3}$ pc;][]{kramer03}  combined with 
the NE2001 model of Galactic electron distribution from \citet{cordes02}, the distance to the pulsar is estimated to be 4 kpc.
However, the estimation of the distance using such methods in this specific region of the sky can be erroneous \citep[see][]{mitra01}
due to the structured and nearby emission of the Gum nebula, a $35\degr$ large nebula located at $\sim$500 pc.
The electron density inside the Gum nebula can vary from 0.1 to 100 cm$^{-3}$ \citep{reynolds76,wallerstein80,dubner92,reynoso97} and
 therefore so can the DM of a pulsar lying in the background of the nebula. 
This complex region is modeled as a simple sphere (the Gum nebula) of homogenous density in the NE2001 model.

In order to better constrain the distance to the pulsar, and search for a potential X-ray nebula, 
an  \textit{XMM-Newton} observation was carried out whose results are presented in this article.
In Sect. \ref{obs}, we present  a detailed morphological and spectral analysis of the pulsar and of its nebula.
A discussion of the distance of the PSR and of the SNR RX J0852.0$-$4622  is presented in Sect. \ref{distance}.
To conclude, a summary of the proposed scenario, the potential relation between PSR\,J0855$-$4644 and the RX J0852.0$-$4622 SNR and the expected TeV emission from the PWN 
 is presented in Sect. \ref{discussion}.

\begin{figure*}[t]
 \centering
   \begin{tabular}{cc}
 {\includegraphics[bb= 0 0 565 685,clip,width=8cm]{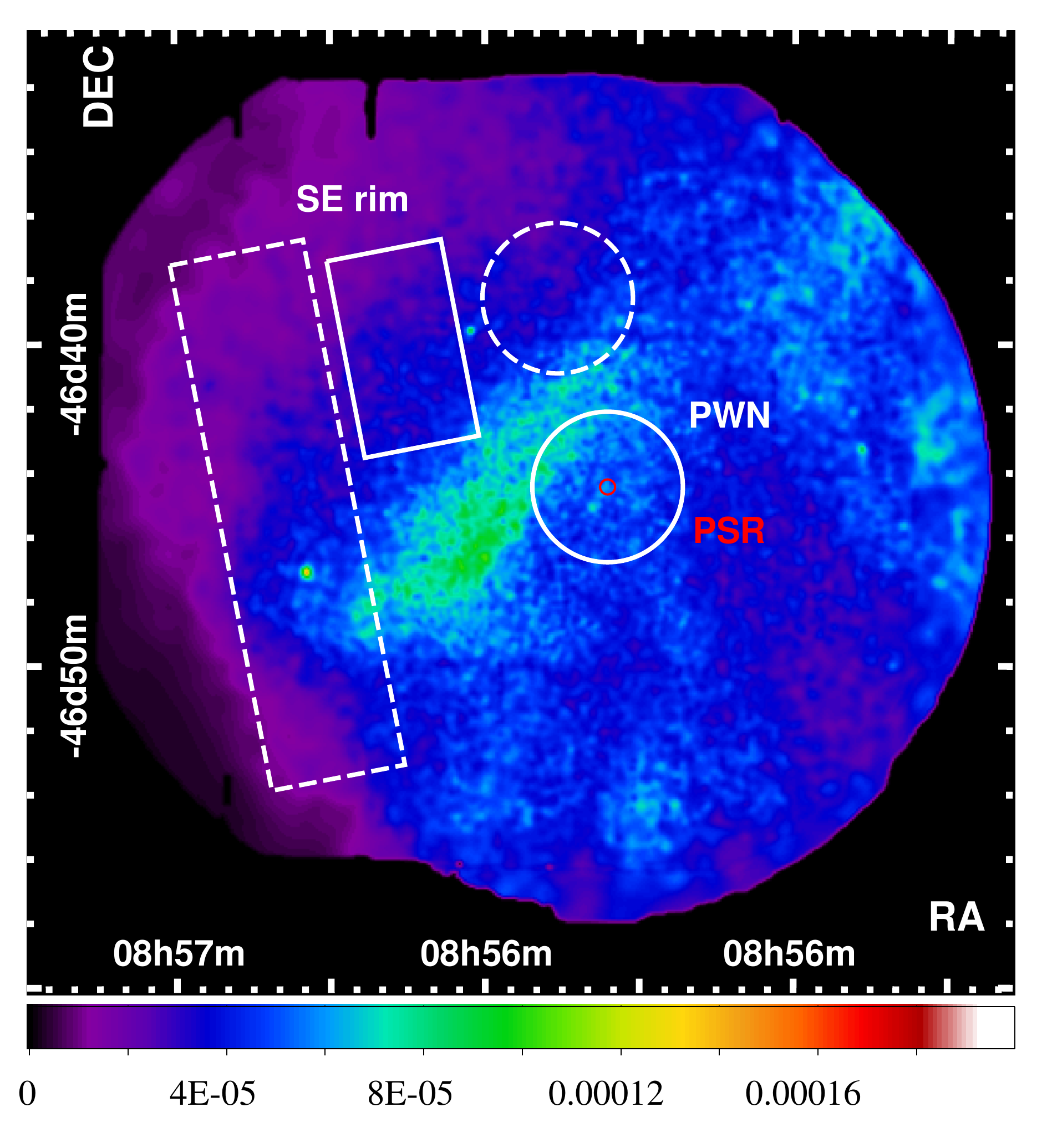} } & \hspace{10mm}
 {\includegraphics[bb= 0 0 565 685,clip,width=7.8cm]{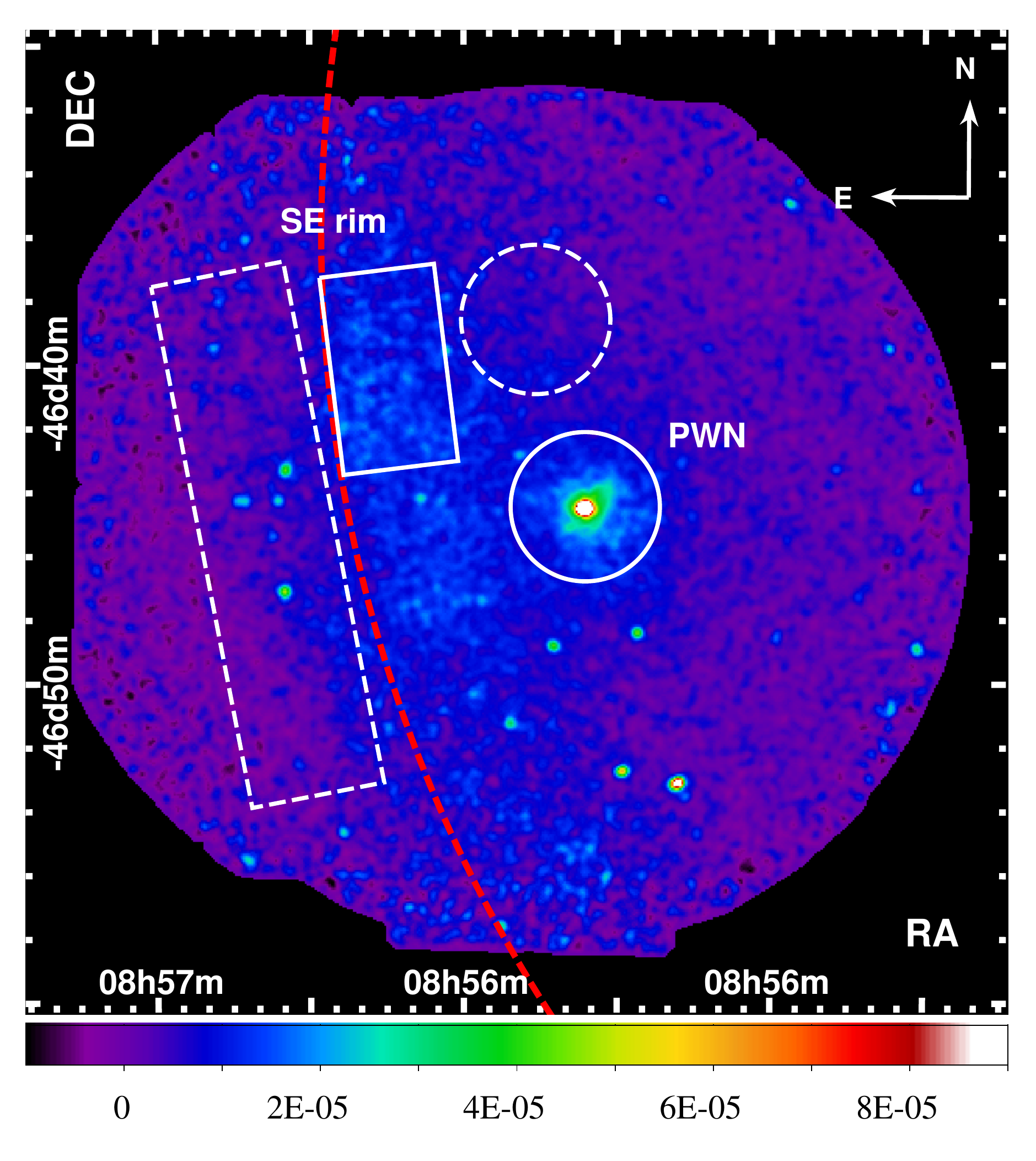} } 

   \end{tabular}

\vspace{-4mm}

\caption{
\footnotesize
\textit{XMM-Newton} background subtracted flux images combining MOS and PN cameras in units of ph/cm$^{2}$/s/arcmin$^{2}$. \textit{Left}: low energy band image (0.5-0.8 keV) 
adaptively smoothed to a signal-to-noise ratio of 10 using the task  \textit{asmooth} provided in the \textit{XMM-Newton} software SAS.
 The position of PSR\,J0855$-$4644 is shown with a red circle.  \textit{Right}: image in the high energy band (1.2-6 keV) smoothed with a Gaussian width of 7''.
 The dashed red line shows approximately the shell of the SNR RX J0852.0$-$4622.
 In solid line we show the spectral extraction regions used to study the rim of RX J0852.0$-$4622  and the PWN whereas the dashed line regions are 
 used to estimate the spectrum of the astrophysical background. 
 }
\label{xmm}
\end{figure*}

\section{X-ray observations}
\label{obs}

The SNR RX J0852.0$-$4622  was first discovered in X-rays by the ROSAT satellite in a high energy image (E$>$ 1.3 keV) 
and has revealed the  energy dependent morphology of this complex region. 
Due to its large angular size (nearly 1$\degr$ in radius), the SNR has only been partly observed by the new generation of X-ray telescopes
and, so far, no mosaic, covering the whole remnant, at high angular resolution is available.
Most observations have been focused on the intriguing bright and thin filaments emitting synchrotron emission in
the North-West of the remnant \citep[see e.g.][]{bamba05}. 

The \textit{XMM-Newton} observation presented in this study is centered on the pulsar and also unveils the, previously unobserved, South-eastern shell of the RX J0852.0$-$4622 SNR.
Several other \textit{XMM-Newton} observations in the North-West  and in the South of the SNR were used in this study to compare the spectral properties of the SNR along the shell.

\subsection{Data analysis}

The region of the pulsar was observed with \textit{XMM-Newton} for 55 ks using the Full Frame mode for the three EPIC instruments. 
After flare screening, the remaining exposure time is 40 ks and 29 ks for the MOS and PN cameras respectively.
For the image generation, the instrumental background was derived from a compilation of blank sky observations \citep{cr07} renormalized in the 10-12 keV energy band.
The background subtracted and vignetting corrected images in a low (0.5-0.8 keV) and high energy band (1.2-6 keV) are presented in Fig. \ref{xmm}.
Those images clearly show two very different  facets of the Vela region. While the low energy image is dominated by the large scale structure of  thermal emission
from the Vela SNR, the high energy band reveals a faint rim of the RX J0852.0$-$4622 SNR and the X-ray counterpart of the pulsar.
Interestingly, diffuse emission is also seen surrounding the pulsar. The spectral properties of the pulsar and of the diffuse emission are presented in the following sections. 

The thermal emission from the Vela SNR is extremely bright at low energy (E$<$1 keV) and therefore the statistics gathered with the PN camera is very high. 
We have observed a calibration issue with the spectral response of the PN camera, particularly visible in the Oxygen lines.
This is a known issue for the \textit{XMM-Newton} team and following their recommendation an offset in energy in the PN spectral response function (\textit{rmf}) was introduced via the \textit{gain} function in Xspec.
All spectra were fitted using MOS1+2 and PN.  For graphical purposes, only the MOS1+2 spectra are shown along with the best-fit obtained with the three instruments. See Appendix \ref{sect:appendix} for more details.

\subsection{Morphology of the nebula}

\begin{figure}[t]
 \begin{tabular}{c}

{\includegraphics[bb=-90 0 572 692,clip,width=6.7cm]{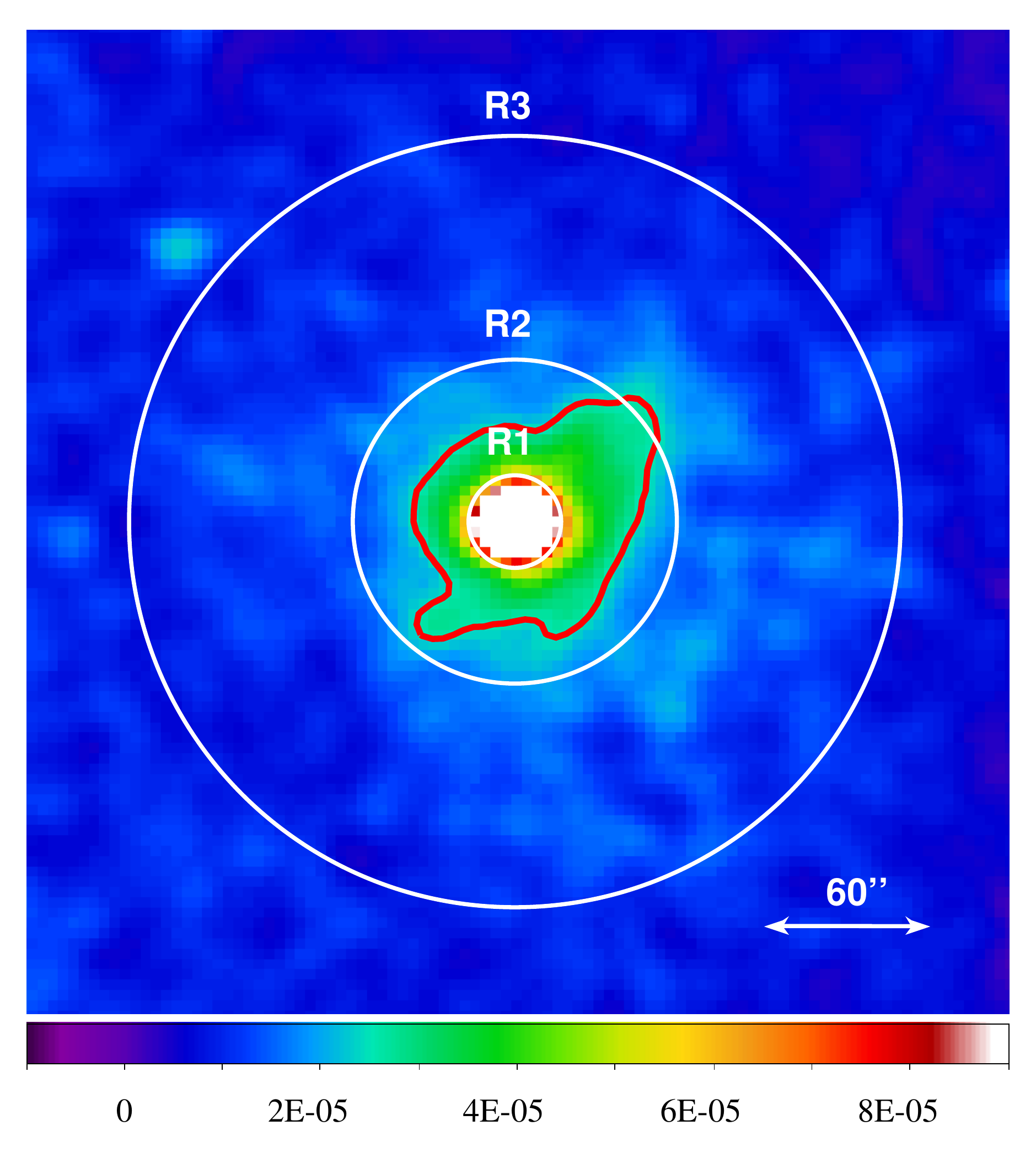} } \vspace{-0.4cm}   \\
{\includegraphics[width=8.cm]{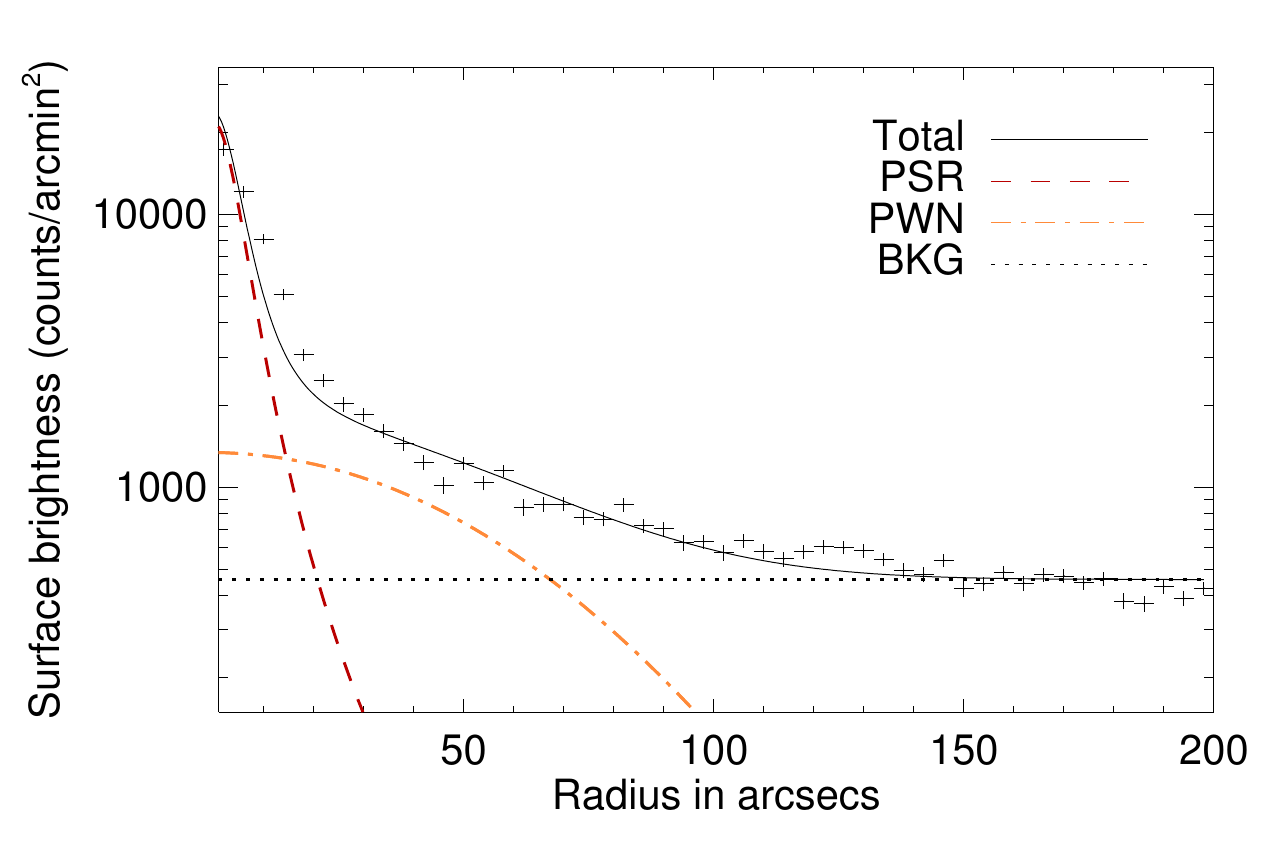} } \vspace{-0.4cm} \\
{\includegraphics[width=8.cm]{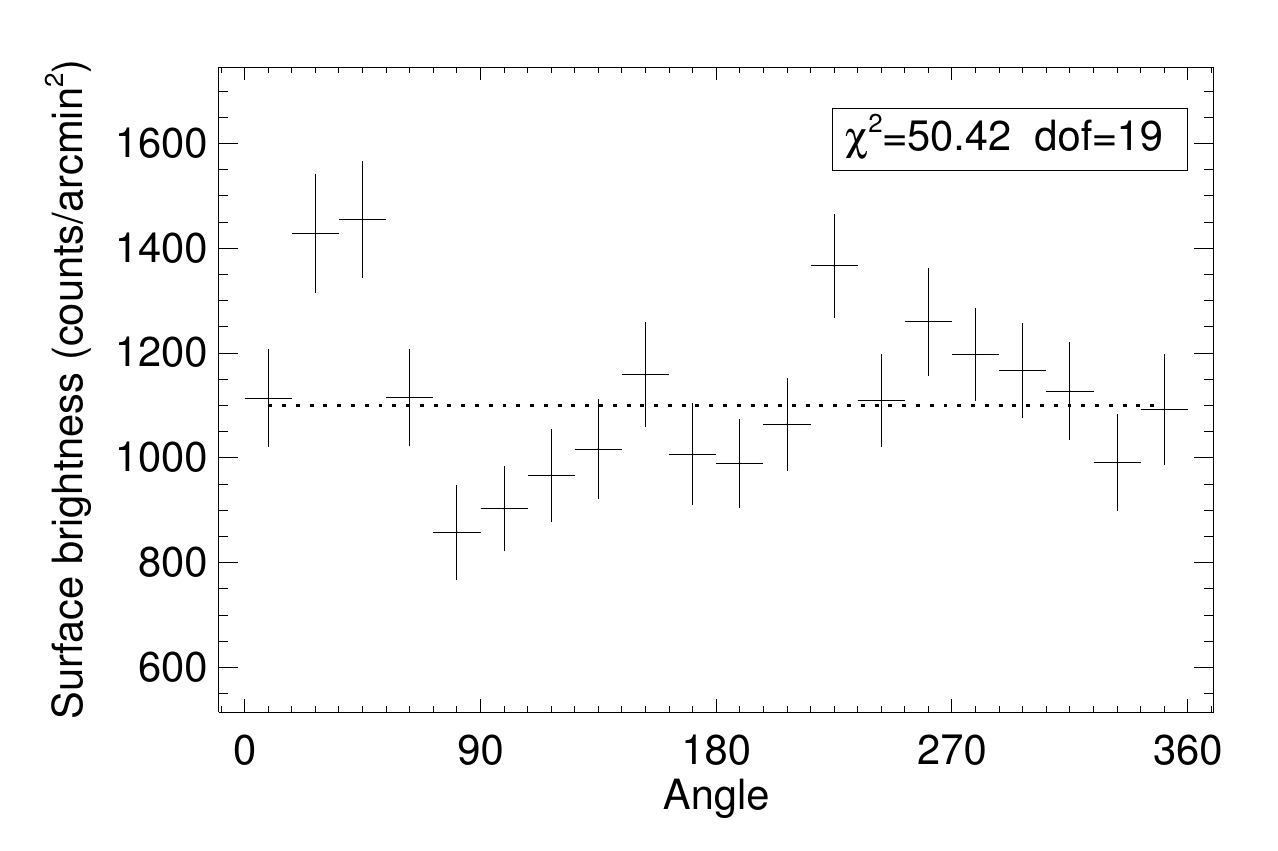} } \\

\end{tabular}

\vspace{-0.4cm}
\caption{
\footnotesize
 \textit{Top}: Zoom on the PWN region in the 1.2-6 keV image shown in Fig. \ref{xmm} (right).  North is up and East to the left. The region is annotated with the annulus regions of radii R1, R2 and R3 (15'', 60'' and 150'' respectively) used in the morphological and spectral analysis. A contour at the level of $2.5\times10^{-5}$ ph/cm$^{2}$/s/arcmin$^{2}$ is overlaid in red to outline the  morphology of the inner nebula.
 \textit{Middle}:  radial profile of the PWN in the high energy band (1.2-6 keV) extracted from a region centered on the pulsar.
 The best-fit two components model (point-like + diffuse) is overlaid.
 \textit{Bottom}: 
 Azimuthal profile extracted from the annulus region (15''$<$ r $<$60'') illustrated in Fig. \ref{xmm}  (right)   in the high energy band. Position angle 
0$\degr$ corresponds to the West and 90$\degr$ to the North. Two significant enhancements over the mean value (dashed line) are seen at  $\sim$50$\degr$ and
$\sim$230$\degr$. The fact that  these knots of emission are separated by 180$\degr$ is suggestive of a jet morphology. 
}
\label{profiles}
\end{figure}

To further investigate the morphology of this diffuse emission, a radial profile (shown in Fig. \ref{profiles}, middle), 
centered on the position of the pulsar, was extracted in the high energy band (1.2-6 keV).
The profile was then fitted with a combination of a point-like component (using the \textit{XMM-Newton} PSF from the calibration files),
  an extended component (a Gaussian profile) and a constant respectively  representing the pulsar,  the nebula and the background.
The best-fit Gaussian width deconvolved from the PSF of the extended component is 45"$\pm$1.5'' (unless otherwise noted, all error bars are given at a 90\% confidence level)
 and the diffuse emission around the pulsar is  detected up to $\sim$150''.
 The large scale emission is fairly symmetric and no indications of a trail or of a bow-shock structure 
 (reminiscent of a fast moving pulsar) are observed.

Interestingly, at smaller scale the inner part of the nebula (r $<$ 60'') seems to show some azimuthal variation. To test the significance of these
variations, an azimuthal profile was extracted from an annulus region (R1=15''$<$ r $<$R2=60'', see Fig. \ref{profiles} top panel for region definition).
The resulting brightness distribution presented in Fig. \ref{profiles} (bottom) is not compatible with a flat profile and exhibits features,
 180$\degr$ apart,  at 50$\degr$ and 230$\degr$  which are suggestive of jet-like structures as observed in many other young pulsars
 \citep[see ][]{kargaltsev08}. However, the limited spatial resolution of \textit{XMM-Newton} hampers a study of the immediate environment of 
the pulsar.  To confirm the presence of jets in the nebula and to search for a torus, high spatial resolution observation with the 
\textit{Chandra} X-ray satellite would be needed.

\subsection{Spectral analysis of the nebula}
\label{specanal}

To extract the spectral properties of the nebula, the instrumental and astrophysical backgrounds have to be modeled carefully in this 
complicated region of the sky.
The instrumental and particle induced background were derived from a compilation of observations with the filter wheel 
in closed position\footnote{\tiny http://xmm2.esac.esa.int/external/xmm\_sw\_cal/background/filter\_closed/} and renormalized in the 10-12 keV energy band over the whole  field of view. The spectral data were fitted using Cash statistics  and 
 the goodness-of-fit was assessed  trough Monte Carlo simulations using the  \textit{goodness} command in Xspec (v12.5).
 This command returns the fraction of simulated spectra whose fit statistics value is less than the one obtained on the real data which is 
 equivalent to 1-$p$-$value$\footnote{ The $p$-$value$ is the probability of obtaining a fit statistic greater than the observed value.}.
 If the model is a good representation of the data, a $p$-$value$ around 0.5 is expected. If the $p$-$value$ value is close to 0, then the model is clearly a bad representation of the data.
The Tuebingen-Boulder ISM absorption model (\textit{tbabs}) is used for all absorbed components together with the abundances of \citet{wilms00}.

The astrophysical background in this region is particularly complex as illustrated in Fig. \ref{xmm}.
 The low energy part of the spectrum is  dominated by the thermal emission from the 
Vela SNR and from the Local Bubble while in the high energy band non-thermal emission from RX J0852.0$-$4622 (the contribution from the SNR's rim
at the PWN's position is not negligible; see Sect. \ref{rimprof}) and the cosmic X-ray background (CXB) are present.
To model this background a spectrum from a region (dashed circle in Fig. \ref{xmm}, right) located at the same distance from the 
shock front of the RX J0852.0$-$4622 SNR (to have similar spectral properties) has been obtained.

The spectrum from this background region is well fitted by a two component model ($p$-$value$=0.24). The first component is  
an  absorbed VAPEC (Variable abundances astrophysical plasma emission code) model
which represents the thermal contribution from Vela.
The X-ray absorption column $N_{\rm H}$ of this component is not well 
constrained with \textit{XMM-Newton}'s energy range (0.4-10 keV) and is therefore fixed to $5\times10^{20}$ cm$^{-2}$, the typical value derived by \citet{aschenbach95} 
using data from the ROSAT satellite whose energy range extends to lower energy  (0.1-2.4 keV). 
The best-fit parameters for this thermal component are the following: $kT$=0.14 keV, Oxygen abundance=0.45\footnote{The abundance of Neon was also let free as a test
 but was found to be compatible with solar abundance and was therefore fixed to solar value in the model.} and a normalization of $7.1\times10^{-3}$ cm$^{-5}$.
The second component is an absorbed power-law model representing the contributions from RX J0852.0$-$4622
and the CXB. Adding another component to represent separately the CXB, with spectral parameters fixed from \citet{deluca04},
did not improve the quality of the fit. The best-fit parameters give  $N_{\rm H}$=$(0.45 \pm 0.10)\times  10^{22}$ cm$^{-2}$, $\Gamma$=1.90$\pm$0.04 and
$norm$=$(2.70 \pm 0.11 )\times10^{-4}$ keV$^{-1}$cm$^{-2}$ s$^{-1}$ at 1 keV.

The best-fit model obtained from the background spectrum is then used as a fixed template model in the spectrum of the nebula 
(annulus: R1=15''$<$ r $<$R3=150'') where the normalization is 
set to the ratio of geometrical areas for the non-thermal emission and let free for the thermal emission as the emission from the Vela SNR
is seen to slightly vary in the low energy map shown in Fig. \ref{xmm}. 
All the other parameters ($kT$, power-law absorption column and spectral index) for the background model are kept fixed.
To represent the emission from the nebula, an absorbed power-law model was added to the previously derived background model. 
The spectrum from the nebula is well modeled (see Fig. \ref{spec}, top) with the sum of the background + power-law model
 ($p$-$value$=0.14) thus confirming the non-thermal nature of the nebula's X-ray emission.

\subsection{Spectral analysis of PSR\,J0855$-$4644}

\begin{figure}[t]

{\includegraphics[bb= -10 -10 442 705,clip,angle=-90,width=8.5cm]{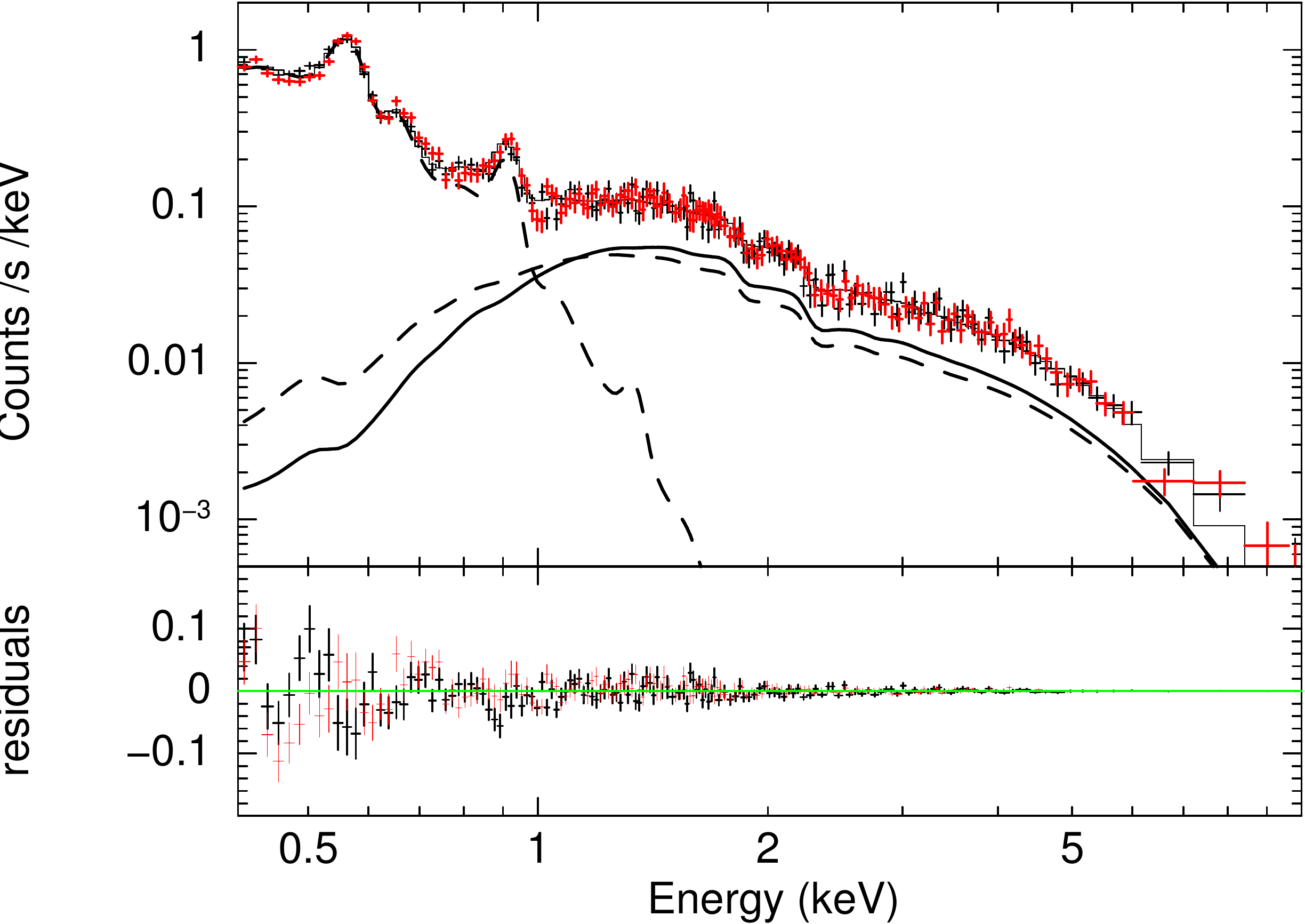}}
{\includegraphics[bb= -10 -10 515 705,clip,angle=-90,width=8.5cm]{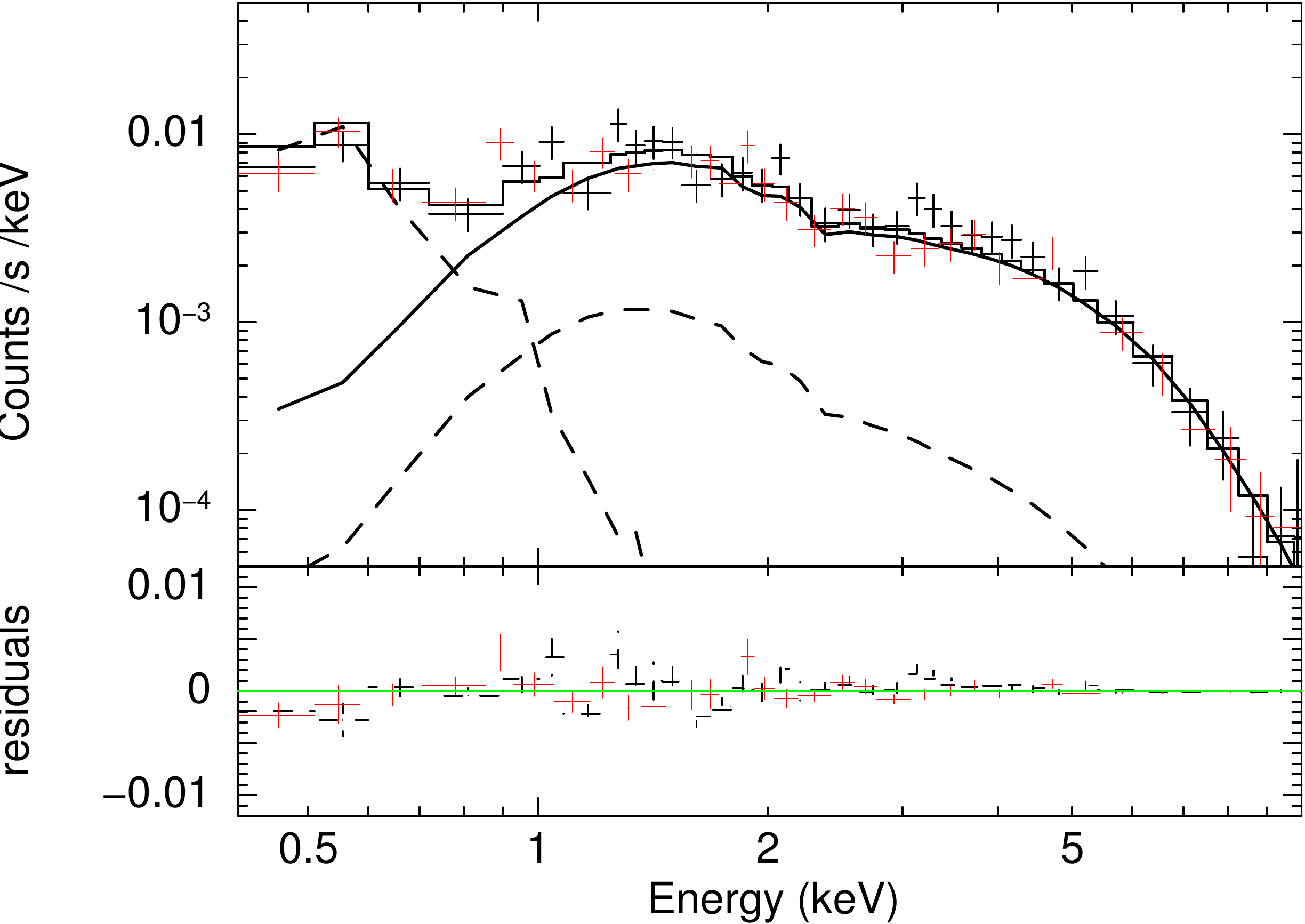}}

\vspace{-0.3cm}
\caption{
\footnotesize
\textit{Top}: X-ray spectrum of the nebula extracted from a 15-150'' annulus region shown in Fig. \ref{xmm}.
The astrophysical background components (dominated by: thermal emission from Vela + non-thermal emission from RX J0852.0$-$4622) are depicted by dashed
 lines while the non-thermal emission of the nebula is illustrated by the solid line.    
\textit{Bottom}:   
X-ray spectrum of PSR\,J0855$-$4644 extracted from a 15'' radius region.
The astrophysical background (dashed lines) is composed of Vela's thermal emission at low energies
 while the contribution of the nebula and of RX J0852.0$-$4622 at higher energies is not negligible. 
 For graphical purposes only the MOS1 and MOS2 data are shown although the PN data were also used in the fitting process.  }
\label{spec}
\end{figure}

In order to derive the X-ray  properties of the pulsar, and in particular the column density along the line of sight to discuss its distance, we have extracted a
 spectrum in a 15'' radius region centered on the pulsar. For this region, the astrophysical background is composed of Vela and the Local Bubble at low energies
 and of the sum of the contributions from the nebula (which is not negligible based on the radial profile modeling in Fig. \ref{profiles}), RX J0852.0$-$4622 and the CXB. 

The astrophysical background  model was obtained from the best-fit model of the spectrum extracted in the PWN region  (15''$<$ r $<$150'').
The non-thermal component of the template was renormalized using the nebula's geometrical model shown in Fig. \ref{profiles} 
(middle) while the thermal component was renormalized using the geometrical ratio of the areas
 (i.e. thermal emission is assumed to be homogeneous at the scale of the PWN).

The spectrum of  the pulsar (presented in Fig. \ref{spec}, Bottom) is well described ($p$-$value$=0.46) at low energies by the fixed template
 (absorbed VAPEC model) and at high energies with two absorbed power-laws one representing the astrophysical background and one for
 the non-thermal emission of the pulsar. The spectral parameters of the pulsar best-fit model are listed 
 in Table \ref{tab}.  A black body model has also been tested to model the X-ray emission of the pulsar but resulted in a poor fit to the data ($p$-$value$=4$\times 10^{-3}$).

The derivation of the $N_{\rm H}$ from the pulsar is sensitive to the normalization of the thermal component, dominating at low energies. 
To estimate the accuracy to which we can fix the normalization of the thermal model, we calculated 
the ratio of brightnesses in the pulsar region over the PWN region in the 0.4-0.6 keV energy band. The resulting ratio is 1.13$\pm$0.08 confirming that the assumption of homogeneous thermal emission at the scale of the PWN 
 is reasonable. When the normalization of the thermal component in the pulsar spectral model is increased by a factor 1.13, the best-fit  value of $N_{\rm H}$  increases to 0.71$^{+0.13}_{-0.11} \times10^{22}$ cm$^{-2}$, within statistical errors of the value quoted in Table \ref{tab}. This test provides an estimate of the systematic error associated with the 
 normalization of the thermal emission at low energies.

\begin{table*}

\centering
\caption{Best-fit parameters obtained on the nebula, the pulsar and different parts of the rim of the RX J0852.0$-$4622 SNR.
For each spectrum, a two components (absorbed VAPEC + absorbed power-law, see Sect. \ref{specanal} for details) astrophysical background model was estimated from the background regions
shown in Fig. \ref{xmm} and \ref{otherxmm}. 
Only the parameters from the non-thermal components are listed in this Table while the parameters of the thermal component are rather constant from one region to
another ($kT=0.14-0.17$ keV and the  $N_{\rm H}$ was fixed to $5\times10^{20}$ cm$^{-2}$ for all the regions).
The unabsorbed non-thermal fluxes of the PWN, PSR and SNR's rim are given in the 2-10 keV energy range and the $N_{\rm H}$ in units of $10^{22}$ cm$^{-2}$. 
The area of the extraction region and the annulus radii for the PSR/PWN analysis are also given.
The error bars are given at 90\% confidence level.}
\label{tab}

\begin{tabular}{l |  c c c c c}     % 8 columns %
\hline

    & \multicolumn{3}{c}{Power-law best-fit parameters}   &  Region area & Annulus radii \\
   & $N_{\rm H} $ & Index & Flux (10$^{-12}$ ergs/cm$^{2}$/s) & arcmin$^{2}$ & arcsec \\

\hline

   \rule[-7pt]{0pt}{17pt} 
	\textit{Pulsar \& PWN}  & & & \\

   \rule[-7pt]{0pt}{17pt} 
   PSR\,J0855$-$4644 &   0.64$^{+0.13}_{-0.11}$ & 1.24$^{+0.09}_{-0.10}$ & 0.26$^{+0.03}_{-0.05}$ & 0.2  & 0-15 \\  

  \rule[-7pt]{0pt}{17pt} 
   PWN &  0.76$^{+0.06}_{-0.05}$ & 2.03$^{+0.06}_{-0.04}$ & 0.88$^{+0.04}_{-0.08}$   & 19.4 & 15-150 \\ 

   \rule[-7pt]{0pt}{17pt} 
  
   Inner PWN &   0.64  & 1.70$^{+0.07}_{-0.06}$ & 0.39$^{+0.02}_{-0.06}$  & 2.9 & 15-60 \\ 

   \rule[-7pt]{0pt}{17pt} 
   Outer PWN & 0.64 & 2.01$^{+0.04}_{-0.04}$ & 0.55$^{+0.03}_{-0.06}$  &  16.5  & 60-150\\  
   
   \hline

      \rule[-7pt]{0pt}{17pt} 
	\textit{SNR}  & & & \\
	
   \rule[-7pt]{0pt}{17pt} 
   South-East rim  & 0.88$^{+0.08}_{-0.06}$ & 2.32$^{+0.09}_{-0.08}$ & 1.83$^{+0.08}_{-0.09}$  & 25.7 &\\   
   
     \rule[-7pt]{0pt}{17pt} 
   South rim &   0.85$^{+0.05}_{-0.05}$ & 2.66$^{+0.05}_{-0.07}$ &   2.88$^{+0.09}_{-0.08}$ & 42.1 & \\ 
   
     \rule[-7pt]{0pt}{17pt} 
   North-West rim & 0.68$^{+0.04}_{-0.05}$ & 2.53$^{+0.05}_{-0.04}$ & 4.10$^{+0.11}_{-0.13}$ & 15.2 & \\ 
   
\hline
\end{tabular}

\end{table*}

To test for a possible spectral cooling in the nebula, 
a spectrum was derived in the inner part  (15''$<$ r $<$60'') and the outer part of the nebula (60''$<$ r $<$ 150'').
For both regions, the astrophysical background was fixed to the best-fit model obtained from the whole PWN in Sect. \ref{specanal}.
In order to limit the degeneracy between the absorption column and the photon index, the $N_{\rm H}$ of non-thermal emission from the nebula
 was fixed to the value derived from the pulsar.

The best-fit parameters  of the inner and outer region of the nebula are presented in Table \ref{tab}.
 A significant steepening of the spectral index towards the external region of the nebula is observed
(from 1.70$^{+0.07}_{-0.06}$ to 2.01$^{+0.04}_{-0.04}$). This is consistent with the scenario of a population of electrons cooling
 as they are advected away from the pulsar.

 In the view of the spectral parameter evolution in the inner nebula, the background model for the pulsar was revisited.
The inner region of the nebula, instead of the whole nebula, was tested as the template background model for the pulsar.
 The best fit parameters of the pulsar with this new background model are: 
 $N_{\rm H}$=0.59$^{+0.14}_{-0.13} \times10^{22}$ cm$^{-2}$ and Index=1.20$^{+0.08}_{-0.11}$.
 When rescaling the normalization of the thermal component  by the same 1.13 factor (as tested in the previous section), 
 the  $N_{\rm H}$ slightly increases to 0.67$^{+0.15}_{-0.13} \times10^{22}$ cm$^{-2}$ and Index=1.23$^{+0.10}_{-0.11}$.
 Those parameters are compatible with the ones presented in Table \ref{tab}
 and in the following discussion only the best fit parameters from the pulsar presented in the table will be used.

 \subsection{Spectral analysis of the rim of the SNR}
\label{rims}

\begin{figure}[]
%{\includegraphics[bb= 0 0 565 500,clip,width=8.5cm]{20799f8.pdf} }
%{\includegraphics[bb= 0 0 565 500,clip,width=8.5cm]{20799f9.pdf} }
{\includegraphics[bb= -25 0 505 435,clip,width=8.7cm]{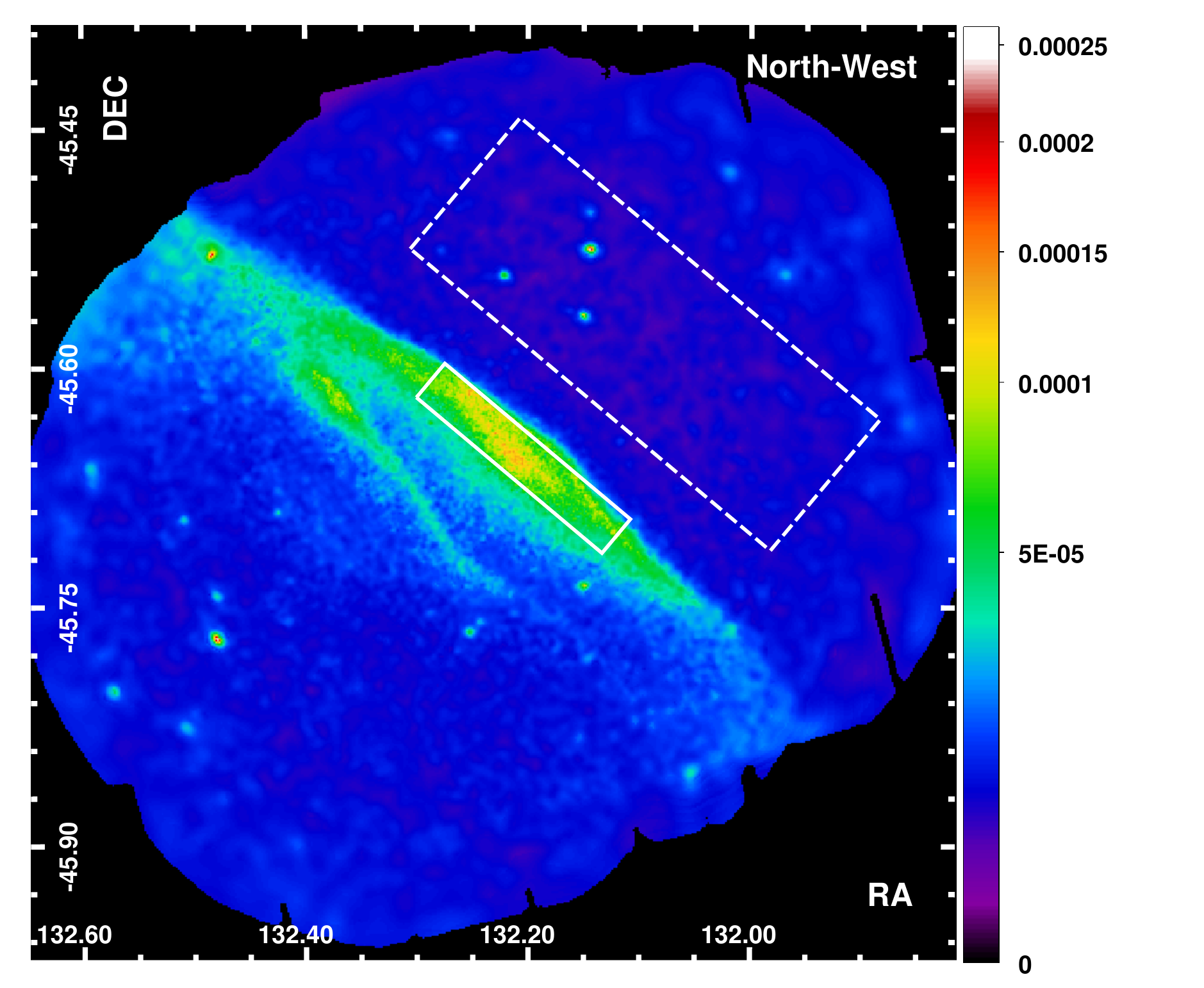} }
{\includegraphics[bb= -25 0 505 435,clip,width=8.7cm]{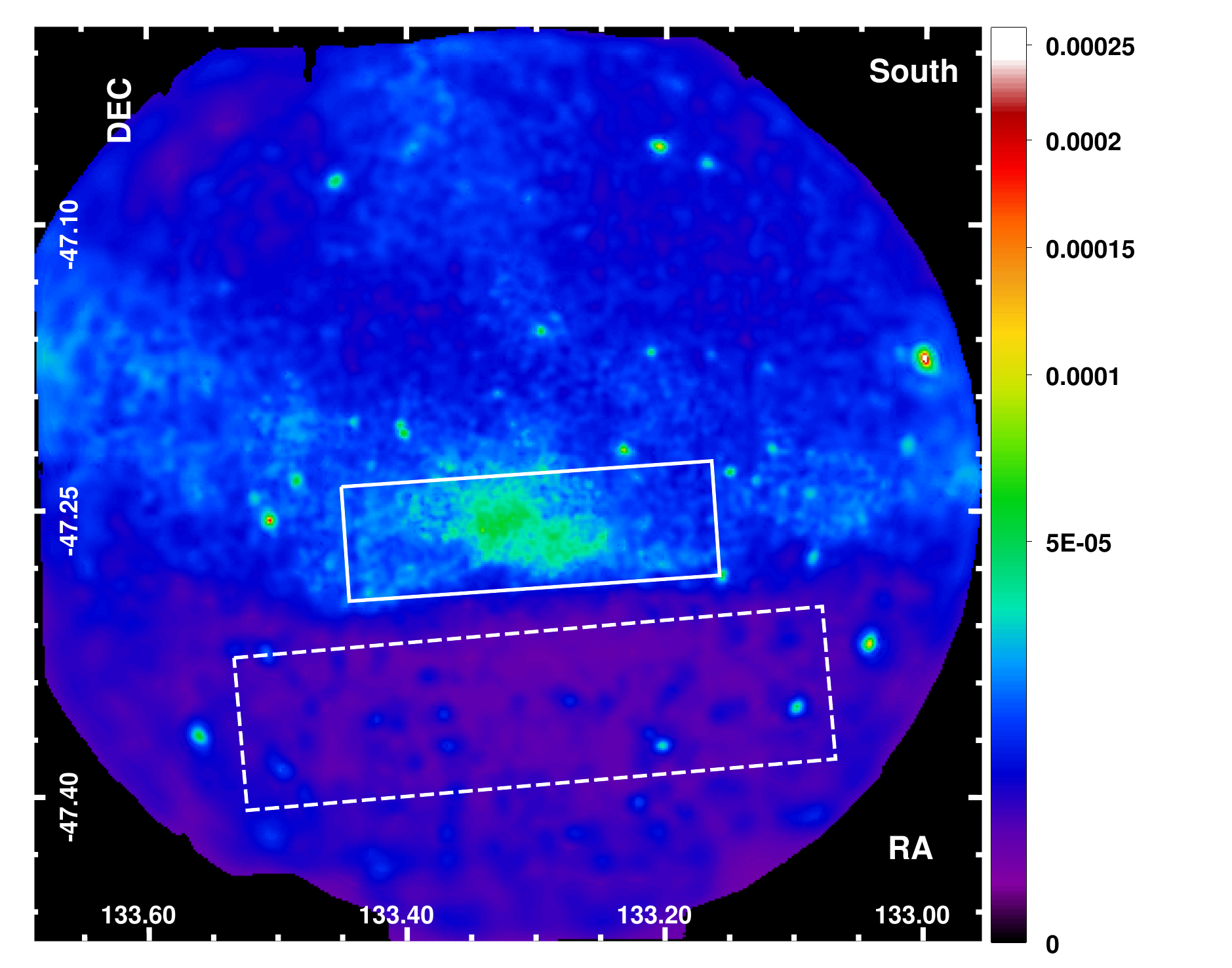} }

\vspace{-0.2cm}
\caption{
\footnotesize
 \textit{XMM-Newton} background subtracted flux images (combining MOS and PN cameras) of the North-West (Top) and South (Bottom) pointings of RX J0852.0$-$4622 in the 1.2-6 keV energy band.
 As in Fig. \ref{xmm}, the solid lines illustrate the regions of spectral extraction whereas the dashed lines represent the regions used to
  estimate the astrophysical background. 
 The images have been  adaptively smoothed in order to reach a S/N of 10, the color scale is in square root 
  and the units are ph/cm$^{2}$/s/arcmin$^{2}$.}
\label{otherxmm}
\end{figure}

To discuss the relative distances of the SNR and of the PWN, we carried out a spectral analysis  of different regions of the RX J0852.0$-$4622 SNR.
In the same pointing where the PWN is observed, we extracted a spectrum from the rim in the region labeled \textit{SE rim}  in Fig. \ref{xmm}. 
The region does not fully encompass the rim in order to avoid the bright structure of thermal emission from Vela that would hamper a proper determination of the absorption column.
As in the previous sections, an astrophysical background model (absorbed VAPEC + absorbed power-law) is fitted to a region outside the SNR  represented by the dashed box in Fig. \ref{xmm}.
The total spectral model for the SNR's rim is the sum of the previously estimated background model plus an absorbed power-law representing the synchrotron emission from the SNR.
The parameters of the background model are fixed in the analysis of the rim except the normalization of the thermal emission.
 The best-fit parameters of the non-thermal emission from the rim are listed in Table \ref{tab}. The same procedure is then applied to other regions of the SNR where X-ray observations were available namely the North-West and the South of the SNR.
The 1.2-6 keV image of those regions together with the spectral extraction regions and background regions are shown in Fig. \ref{otherxmm}.
The eastern rim of the SNR has also been observed by \textit{XMM-Newton} but the observation has been, unfortunately, strongly affected by solar flares and only 5 ks out of 36 ks of observation are usable.
As a consequence, no reliable spectral information could be derived from the diffuse and faint emission from the SNR in this region. 

The resulting  best-fit parameters of the spectral analysis from three different regions of the SNR are listed in Table \ref{tab}.

\subsection{Radial profiles of the SNR's rim}
\label{rimprof}

\begin{figure}[t]
\includegraphics[width=9cm]{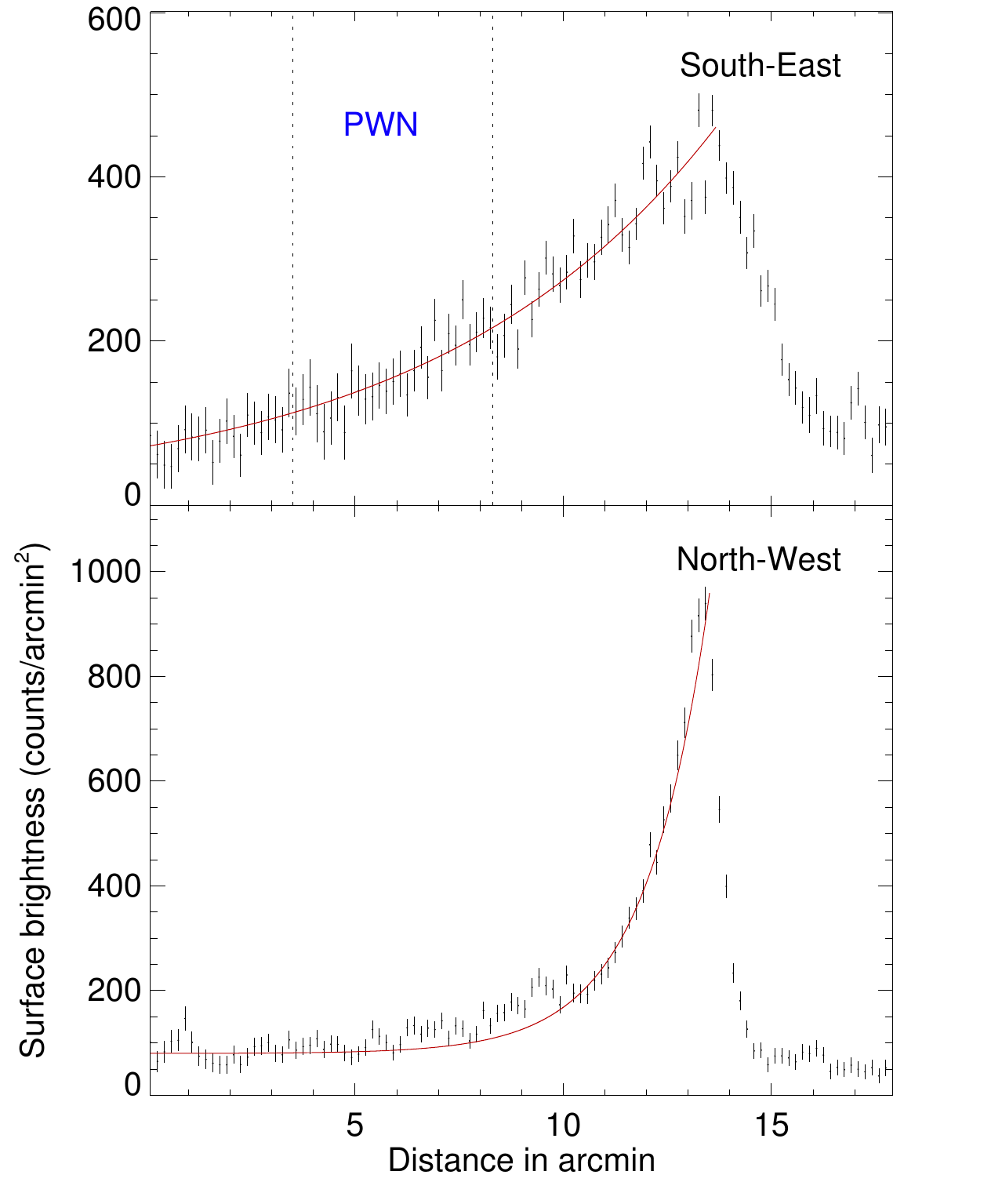}

\vspace{-0.25cm}
\caption{
\footnotesize
Radial profiles obtained in the rims  of the RX J0852.0$-$4622 SNR in the South-East (top) and the North-West (bottom) along the spectral extraction regions shown in Fig. \ref{xmm},\ref{otherxmm} .
In the South-East, the distance of the PWN from the front shock is represented by the dashed lines.
The best-fit exponentially decreasing model in the downstream side of the shock is illustrated with the solid lines. }
\label{rimradprof}
\end{figure}

To estimate the non-thermal contribution of RX J0852.0$-$4622 at the PWN position, a radial profile was extracted from a region of the same width 
and direction as the region labeled \textit{SE rim} in Fig. \ref{xmm} (right panel) but with a different length. 
The resulting profile, shown in Fig. \ref{rimradprof} (top), demonstrates that the emission of RX J0852.0$-$4622 is not negligible at the reported distance of the 
PWN behind the shock front and its implementation in the astrophysical background modeling, in Sect. \ref{specanal}, is justified.

In order to measure the scale width of this filament, the profile was fitted with an exponentially decreasing model 
of the form $e^{-|r_{0}-r|/w_{d}}$ where $d$ denotes the downstream side of the shock \citep[same geometrical model as in][ where 
the North-West rim of the SNR is studied]{bamba05}.
The best-fit model is overlaid in Fig. \ref{rimradprof} (top) and the best-fit parameters is $w_{d}$= 6.85 $\pm$ 0.55 arcmin.
For comparison, the radial profile of the North-western region is shown in Fig. \ref{rimradprof} (bottom).
The latter profile was obtained from a region of the same direction than the white solid region shown in Fig. \ref{otherxmm} (top).
The same geometrical model was used and the  best-fit parameters are $w_{d}$= 1.61 $\pm$ 0.05 arcmin.
We note that the downstream scale of the rim in this region is nearly four times thinner than in the South-East region.
The implication of those different widths on the value of the magnetic field that can be derived is discussed 
in more details in Sect. \ref{rimwidth}.

\section{Distance estimate}
\label{distance}

 \begin{figure}[]

%{\includegraphics[bb= 55 340 568 870,clip,width=8.5cm]{20799f11.pdf}}
{\includegraphics[width=8.5cm]{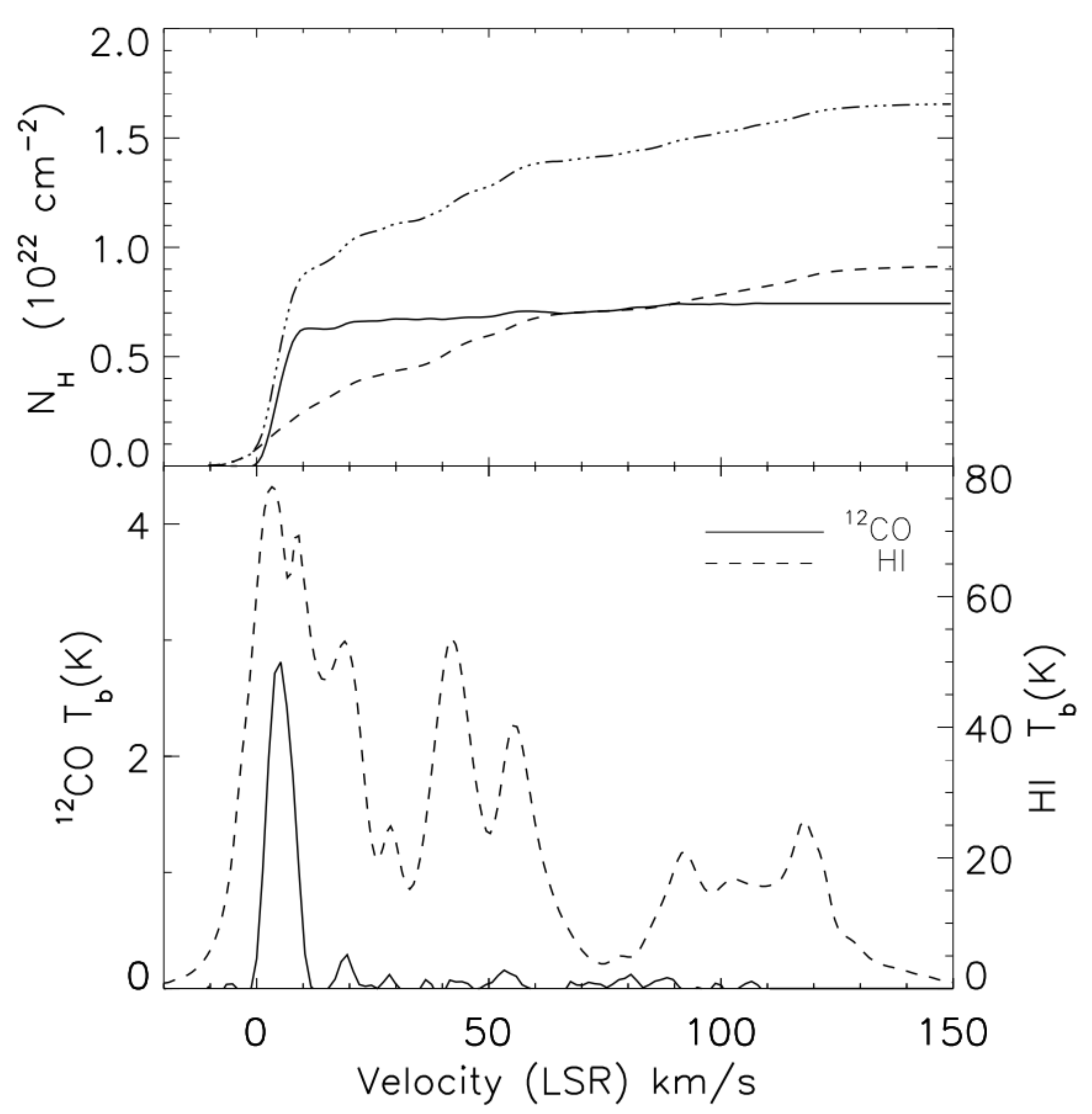}}

\vspace{-0.3cm}
\caption{ 
\textit{Top:} cumulative absorption column density (dot-dashed line) together with the $^{\rm 12}$CO and HI 
contributions as a function of the radial velocity integrated on a 0.1$\degr$ box region centered on the pulsar.
\textit{Bottom:} corresponding $^{\rm 12}$CO and HI spectra at the position of the pulsar.
The Vela molecular ridge, centered at $V_{\rm LSR}$$\sim$5 km/s, is the main $^{\rm 12}$CO structure along the pulsar's line of sight.    }
\label{spectrumgas}
\end{figure}

\begin{figure*}[t]

\includegraphics[width=18cm]{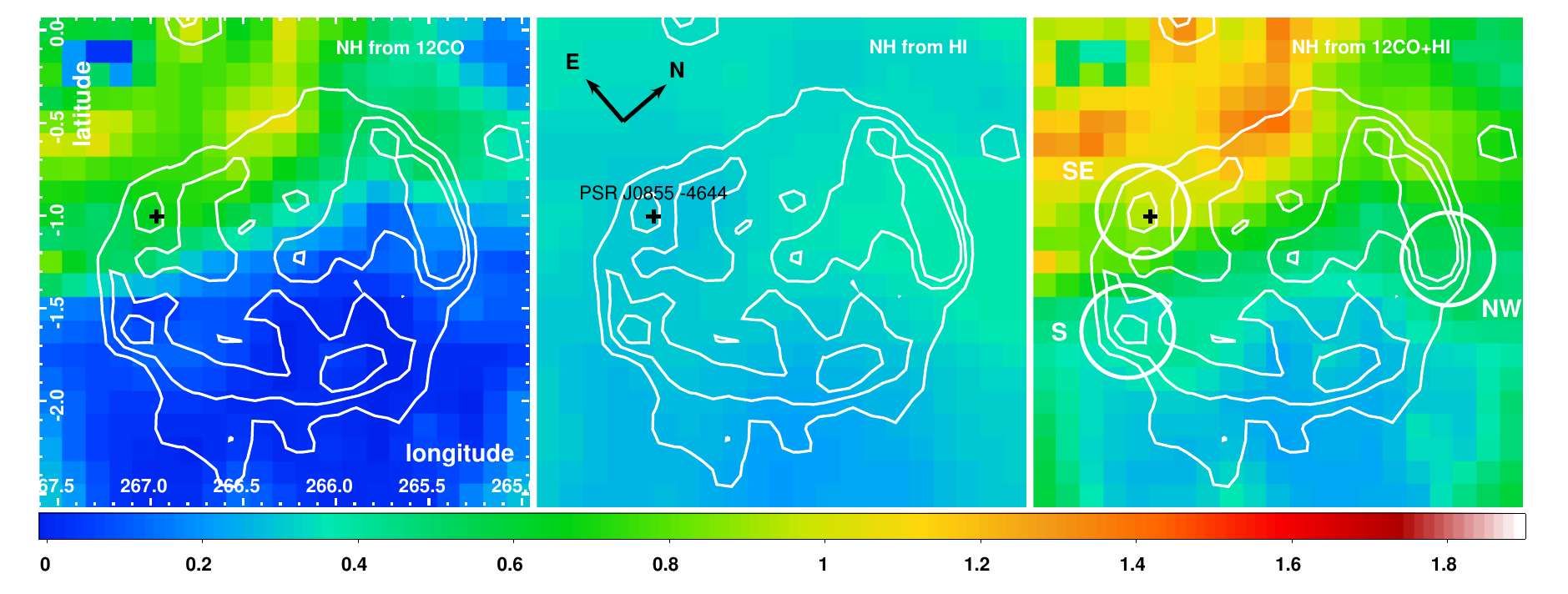}  
\includegraphics[bb= -300 0 450 28,clip,width=12cm]{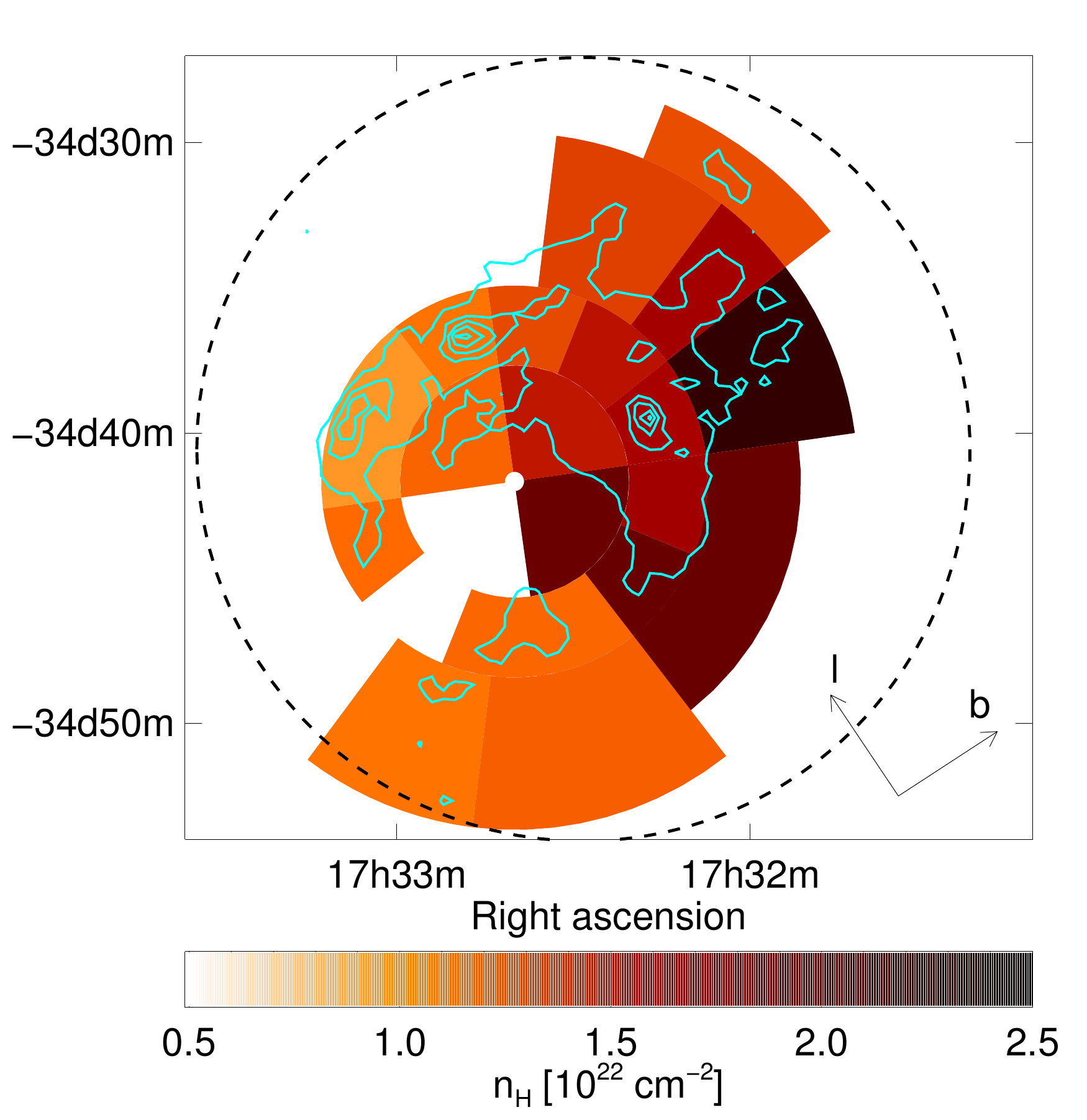}

\vspace{-0.cm}
\caption{
\footnotesize Maps of the absorption column density, derived from the molecular gas (left panel), the atomic gas (middle panel) and the sum of the two previous components  (right panel) when integrating over radial velocities from -10 km/s to 15 km/s (see Sect. \ref{distance} for details). This  specific velocity range provides an estimate of the absorption column for a source located just behind the Vela molecular ridge. The position of PSR\,J0855$-$4644 is shown with a black cross and the \textit{XMM-Newton} pointings used in this study are represented by the white circles in the right panel. The white contours are taken from a ROSAT high energy image (E$>$1.3 keV) and reveal the SNR RX J0852.0$-$4622.  }
\label{NHmap}
\end{figure*}

The method  to  tackle the issue of the distance to the pulsar
 is based on the existence of a strong contrast of integrated $^{12}$CO  between the eastern 
 and the western parts of the SNR, as shown in Fig.  \ref{NHmap} (left).
This contrast is due to the Vela molecular ridge (VMR) cloud C \citep[see ][ for cloud nomenclature]{murphy91},
 lying at a distance of 700$\pm$200 pc.  One of the method to derive the latter distance is an infrared photometric study
  of the star-forming regions in the VMR \citep[see ][ and references therein]{liseau92}. 
  Other independent methods, discussed in \citet{murphy91}, give a similar distance.
  This cloud is the main contributor of $^{12}$CO emission  along the line of sight
 as seen in Fig. \ref{spectrumgas} (bottom panel) and is located at a radial velocity in the local standard of rest of $V_{\rm LSR}\sim5$ km/s.
 If the SNR RX J0852.0$-$4622 lies behind the VMR, the observed gradient in $^{12}$CO should also be visible
 in the measurement of the X-ray absorption column.

To investigate such a correlation, absorption column density maps (presented in Fig. \ref{NHmap})
 are calculated using HI data from the Parkes Galactic all sky survey 
\citep{mcclure09,kalberla10} and $^{12}$CO data from the \citet{dame01} survey.
To fully enclose the gas material lying between the observer and
 the back of the VMR, the data were integrated on radial velocities from -10 km/s to 15 km/s. 
 Due to the intrinsic velocity dispersion of the $^{12}$CO emission line, it is unfortunately not possible to estimate the column density between the observer and the foreground of the VMR. 
The CO-to-H$_{2}$ mass conversion factor and the HI brightness temperature to $N_{\rm H}$ used to produce the maps are respectively of
$1.8\times10^{20}$ cm$^{-2}$ K$^{-1}$ km$^{-1}$ s \citep{dame01} and $1.82\times10^{18}$ cm$^{-2}$ K$^{-1}$ km$^{-1}$ s \citep{dickey90}.
The same conversion factors were used to compute the cumulative absorption column at the pulsar's position shown in Fig.  \ref{spectrumgas} (top panel). 
From this curve, we estimate that, along the line of sight of the pulsar, the $N_{\rm H}$ between the observer and the back of the VMR is 0.97 $\times 10^{22}$ cm$^{-2}$
while the total $N_{\rm H}$ along the line of sight is 1.65 $\times 10^{22}$ cm$^{-2}$.

\begin{figure}[]
%\resizebox{\hsize}{!}{\includegraphics[bb= 45 350 580 750,clip]{20799f14.pdf}}
\resizebox{\hsize}{!}{\includegraphics{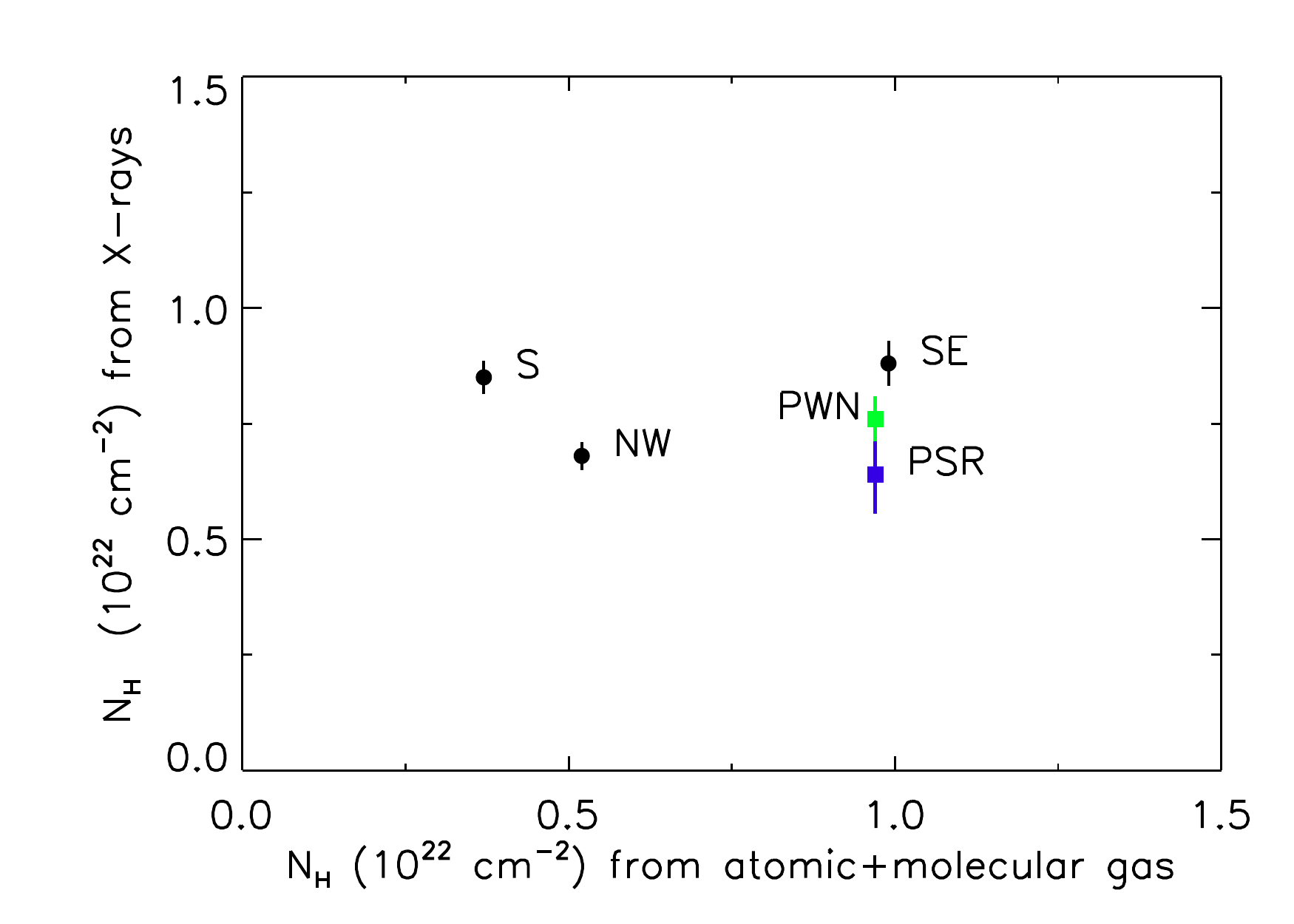}}

\vspace{-0.3cm}
\caption{
\footnotesize
Comparison plot of the absorption column density values derived from the atomic and molecular gas and from the X-rays using \textit{XMM-Newton} observations.
The X-ray columns derived towards the rims of the RX J0852.0$-$4622 SNR,  PSR\,J0855$-$4644 and the PWN are represented by black circles,  a blue square and a green square respectively. The errors on the X-ray $N_{\rm H}$ are 1 $\sigma$ error bars.
The label over each point corresponds to regions where the values have been measured as defined in Fig. \ref{NHmap}.}
\label{correl}
\end{figure}

Using the previously generated maps, a comparison plot is built (see Fig. \ref{correl}) where the $N_{\rm H}$  
derived from the gas maps and integrated up to the VMR 
is plotted against the best-fit values of the X-ray absorption for several positions in the sky.
No correlation is observed in such a plot indicating that the SNR is lying in the foreground of the VMR.
In addition we note that the $N_{\rm H}$ towards the pulsar (0.64$^{+0.13}_{-0.11} \times 10^{22}$ cm$^{-2}$) 
is slightly less than the column towards the SNR's rim in the same region (0.88$^{+0.08}_{-0.06}\times 10^{22}$ cm$^{-2}$)
suggesting that the pulsar is closer than or at the same distance as the SNR RX J0852.0$-$4622.

 An additional test to determine whether the pulsar is in the foreground or background of the VMR
 is to look for the ratio of neutral hydrogen (related to the X-ray absorption column $N_{\rm H}$) to free electrons
 (given by the pulsar radio dispersion measure DM). As discussed in \citet{gaensler04}, this ratio has been observed to be high 
for pulsars which are known to lie behind dense molecular material (e.g. $N_{\rm H}/DM$=85 (40) for PSR\,J1747$-$2958 and 
PSR\,1757$-$24 respectively) whereas the typical value for other pulsars is of the order of 5-10.
 The value we derive for PSR\,J0855$-$4644 is 8$\pm$1.5 which suggests that the pulsar is not located behind dense material.

We therefore conclude that both the SNR and the pulsar are lying in the foreground of the VMR i.e. at a distance d $\leq$ 900 pc.
This distance constraint to the RX J0852.0$-$4622 SNR is compatible with a recent estimate based on the measurement of the proper motion
of the shock in X-rays in the North-western part of the rim by \citet{katsuda08} who derived d$\sim$750 pc.
Concerning the pulsar, a closer distance (than the 4 kpc derived from the radio DM measurements) confirms a suggestion made by \citet{redman05} who derived
a distance of $\sim$750 pc.

\section{Discussion}
\label{discussion}

\subsection{Proposed scenario}

In the previous sections, we have shown that RX J0852.0$-$4622 and  PSR\,J0855$-$4644 are lying in the foreground or at the interface of the VMR.
The SNR and the pulsar are therefore thought to be nearby (d$\leq$ 900 pc) but surprisingly the absorption columns derived from the X-ray analysis show  high values  ranging from  0.64 to 0.88$\times 10^{22}$ cm$^{-2}$. This is intriguing as the $N_{\rm H}$ towards the Vela SNR  whose distance is known precisely\footnote{The distance to the Vela pulsar has been measured by parallax.} to be d=287$^{+19}_{-17}$ pc  \citep{dodson03} is much lower at $\sim$$10^{20}$ cm$^{-2}$.

To understand this apparent discrepancy, we investigated the spatial distribution and the  density of the local gas towards the Vela region.
We note that due to the intrinsic velocity dispersion of the VMR emission both in HI and in $^{12}$CO (see Fig. \ref{spectrumgas}, bottom) and the proximity of the object, it is not possible to disentangle the emission from the VMR from the foreground emission.
 However, the density of the nearby gas (the Local Bubble) can be probed by other methods such as the study of the absorption line features of NaI and CaII towards stars with known distances \citep[see ][]{lallement03,welsh10}. 
These studies revealed interstellar tunnels of low density connecting the Local Bubble to other surroundings cavities. 
One of these tunnels is in the line of sight of the Vela region at $\ell \sim 265 \degr$ extending at least to 250 pc \citep{welsh10}.
This tunnel could explain the low value of the $N_{\rm H}$ measured in X-rays in the direction of the Vela SNR. Using the $N_{\rm H}$ measured towards the Vela SNR ($10^{20}$ cm$^{-2}$), the mean density in this line of sight is of the order of 0.1 cm$^{-3}$.

After the Local Bubble, the line of sight encounters the Gum nebula, a large scale structure centered
at $\ell=262\degr$, $b=-3\degr$ and located at a distance of $\sim500$ pc for a physical radius of $\sim150$ pc and an angular radius of 17\degr. 
Over a distance of 200 or 600 pc (distance between the Vela SNR at 287 pc and the VMR at 700$\pm$200 pc), the required densities must be of the order of 9 to 3 cm$^{-3}$ respectively to reach the  $N_{\rm H}$ value measured in X-rays of $\sim0.6 \times 10^{22}$ cm$^{-2}$ towards PSR\,J0855$-$4644.
When integrating the HI spectrum between the Earth and the background of the VMR ( -10 km/s to 15 km/s), the HI alone contributes to $0.3  \times 10^{22}$ cm$^{-2}$ in the line of sight towards the pulsar corresponding to a mean density of 2-5 cm$^{-3}$ over a distance of 600 and 200 pc respectively.
In addition the HI emission is homogeneously spatially distributed  around the RX J0852.0$-$4622 SNR region  and 
is not correlated with the VMR as shown in the HI map shown  in Fig. \ref{NHmap} (middle panel) thus contributing to all the lines of sight used in the X-ray analysis which is consistent with the relatively high  $N_{\rm H}$ values measured across the SNR. 
The densities aforementioned  are compatible with the density of $\sim$7 cm$^{-3}$ measured in the inner region of the Gum nebula \citep[the thick HI shell discussed in ][]{dubner92,reynoso97} that is spatially overlapping our region of interest. H$_{\alpha}$ observations of the nebula by \citet{reynolds76} 
provide similar values for the densities (1 to 5 cm$^{-3}$). 
One possible origin of the Gum nebula  is that the supershell is produced by the combination of stellar
 winds and repeated SN explosions \citep{mccray87,reynoso97} and it is therefore a plausible location for PSR\,J0855$-$4644 and RX J0852.0$-$4622.

Finally, changing the electron density of the Gum nebula, in the NE2001 model, from 0.5 cm$^{-3}$ (the currently implemented uniform value for the Gum nebula) to 1 cm$^{-3}$  is sufficient  
to reconcile a DM of 239 cm$^{-3}$ pc with a distance to the pulsar of 900 pc.
This does not seem unreasonable with regards to the possible large density variations in the Gum nebula.
 
 In summary, the line of sight towards the pulsar encounters first a tunnel of low density gas ($\sim$0.1 cm$^{-3}$) up to the Vela SNR then 
 enters the Gum nebula thick HI shell with densities of the order of 1-5 cm$^{-3}$. 
 This succession of low and high density environment can explain the low $N_{\rm H}$ towards the Vela SNR and the high $N_{\rm H}$ towards the RX J0852.0$-$4622 SNR and the pulsar. While the HI alone contributes to $0.3  \times 10^{22}$ cm$^{-2}$, additional molecular material in the line of sight could contribute to the measured X-ray absorption either in the foreground of the VMR\footnote{Such material can  unfortunately not be disentangled from the VMR in the $^{12}$CO spectrum.  } or from the external regions of the VMR if the pulsar and SNR
 are lying at the interface of the molecular cloud.

\subsection{Is the pulsar associated with the RX J0852.0$-$4622 SNR ?}
Although the pulsar and the SNR may be at similar distances, an association is unlikely due to the characteristic age of the pulsar 
($\tau_{c}$=140 kyrs) being much larger than the SNR age ($t_{SNR}\sim$ 4000 yrs).
Assuming that $\tau_{c}$ could be very different from the real age,  the required pulsar's kick velocity to
explain its current position would be $\sim$3000 km/s in order to travel 12 pc from the SNR center (1$^{\circ}$ at 750 pc) for a SNR age of 4000 years.
This value is much higher than observed in other pulsars \citep{hobbs05}. Moreover for such a high shock speed a bow-shock or a tail structure should appear in the PWN which is not the case. In addition, a candidate neutron star CXOU J085201.4-461753 \citep[see e.g.: ][]{kargaltsev02} lies near the geometrical center of the SNR RX J0852.0$-$4622 which appears as a more simple scenario.

\subsection{TeV emission from the PWN}

With the new distance derived in this study, the ratio $\dot{E}/d^{2}$ for PSR\,J0855$-$4644 reaches a value  $\ge 10^{36}$ erg/s/kpc$^{2}$ (for d$\le$ 900 pc 
and  $\dot{E}=1.1 \times 10^{36}$ ergs).  According to the TeV $\gamma$-ray PWN population study presented by \citet{carrigan08}, 
such a high $\dot{E}/d^{2}$ pulsar should be detectable  by Cherenkov telescopes in $\gamma$-rays.
The region of the RX J0852.0$-$4622 SNR has been extensively studied by the H.E.S.S. experiment and an excess of TeV $\gamma$-rays 
is observed at the position of the pulsar \citep[see Fig. 3 in ][]{pazarribas12}.
However as the pulsar is in spatial coincidence with the shell of the SNR, it is not possible, given  the spatial 
resolution of the H.E.S.S. telescopes, to disentangle the contribution from the shell of the SNR from the contribution of the PWN.
In addition, the azimuthal profile of the $\gamma$-ray emission along the shell of the SNR does not reveal any significant emission increase
with regards to the surrounding SNR shell emission at the position of pulsar \citep{pazarribas12}.

\subsection{Pulsar contribution to observed e$^{-}$/e$^{+}$ CRs }

This new distance estimate of the pulsar strongly differs from the distance estimated with the radio dispersion
measure (d=4 kpc) and therefore implies that PSR\,J0855$-$4644 is an energetic and nearby pulsar. After the Vela pulsar, it is the most energetic
pulsar within 1 kpc and could therefore  contribute to the spectrum of cosmic-rays e$^{-}$/e$^{+}$  received at Earth  \citep[see][for the contribution of nearby pulsars]{delahaye10}.  Obtaining a good knowledge of the nearby pulsar population is a necessary step to understand the rise in positron fraction with increasing energy measured by the PAMELA experiment \citep{adriani09} and confirmed by the Fermi space telescope \citep{ackermann12}. 

As the pulsar is rather old ($\tau_{\rm c}$= 140 kyrs) and  no traces of the host SNR have been observed, one can reasonably assume that the CR accelerated in the PWN have had time to escape the SNR. The diffusion timescale of those escaping CRs to reach the Earth scales 
as t$_{\rm prop } (E) \approx d^{2}/D(E) $ where $D(E)$ is the diffusion coefficient with a typical value of
 $\sim 3 \times 10^{28} (E/1 \,{\rm GeV})^{\delta}$ cm$^{2}$/s with a large uncertainty on the energy dependence
  $\delta$=0.3-0.6 \citep[see ][ for a review of CR propagation in the Galaxy]{strong07}. 
 Therefore, $t_{\rm prop } (E) \approx 100 \, {\rm kyrs} \, (d/100 {\rm pc}) ^{2} (E/1 {\rm GeV})^{-0.5} $. 
For a typical value of $\delta$=0.5, neither PSR\,J0855$-$4644 nor the Vela pulsar ($\tau_{\rm c}$= 11 kyrs) are thought to contribute to the observed flux of e$^{-}$/e$^{+}$ at 10 GeV 
as the diffusion timescale is respectively 1500 and 280 kyrs (for d=750 and 290 pc respectively).
By equating $t_{\rm prop } =  \tau_{\rm c}$ and for $\delta$=0.5, the energy of the particles that could have diffused up to Earth 
is respectively 1 TeV and 6 TeV for PSR\,J0855$-$4644 and the Vela pulsar.
 
The previous estimates assume that the particles start to diffuse at the birth of the SNR and do not take into account the time for the particles accelerated in the PWN to
break through the shell of the SNR. This escape effect could be not negligible in the case of the Vela pulsar
 \citep[see ][for a study of the diffusive escape effect on the particle spectrum]{hinton11} 
due to the relative young age of the system ($\tau_{\rm c}$= 11 kyrs).

\subsection{ Comparison of the width of the rims of the SNR }
\label{rimwidth}

In the \textit{XMM-Newton} observation of the South-East region of the RX J0852.0$-$4622 SNR, a faint rim of non-thermal emission was detected (see Fig. \ref{xmm}, right).
In Sect. \ref{rimprof}, the downstream thickness of the rim was measured with an exponentially decreasing model.
 We observed that the  filament in the South-East is nearly four times thicker ($w_{d}$= 6.85 $\pm$ 0.55 arcmin) than the bright 
 and thin ($w_{d}$= 1.61 $\pm$ 0.05 arcmin)  filament observed in the North-West of the SNR shown in Fig. \ref{otherxmm}.

Assuming that the thickness of the rim is limited by the synchrotron losses while the electrons are advected away from the shock,
one can derive a lower limit to the value of the downstream magnetic field $B_{d}$. To derive such a value, we adapted the equation 9 from \citet{parizot06}
to keep track of the observed photon energy $E_{\rm ph}$ and obtained the following relation:

\begin{equation}
 B_{\rm d} \geq 37 \,\mu {\rm G} \times V_{\rm sh, 3}^{2/3} \times ( \theta_{\rm obs, arcmin} \times d_{\rm kpc} ) ^{-2/3}  \times  E_{\rm ph, keV}^{-1/3}   
\end{equation}

where  $V_{\rm sh, 3}$ is the shock speed in units of 1000 km/s, $ \theta_{\rm obs, arcmin}$ 
is the observed width of the rim at half maximum in units of arc minutes ($ \theta_{\rm obs}$=$w_{\rm d} \times log(2)$) and $E_{\rm ph}$ the  average emitted photon energy in keV.
For a distance to the RX J0852.0$-$4622 SNR of 750 pc \citep[see ][ and Sect. \ref{distance} of this article for a more detailed discussion of the SNR distance]{katsuda08}, 
a shock speed of 3000 km/s \citep{katsuda08} and a mean photon energy of $\sim$2.5 keV, 
the downstream magnetic field value is  24 $\mu$G and 63 $\mu$G respectively for the South-East and for the North-West rim.

We note that in the North-West region, an X-ray proper motion of the shock ($\dot{\theta}_{\rm sh}$=$V_{\rm sh}/d$) 
has been measured\footnote{The value of $V_{\rm sh}$ and $d$ used in the previous calculation were derived from the aforementioned proper motion measurements.} by \citet{katsuda08}. 
As the ratio of $V_{\rm sh}/d$ can be replaced by $\dot{\theta}_{\rm sh}$ in Eq. (1), 
the estimate of the magnetic field, at least in the North-West, is in fact independent of the hypothesis on the shock speed and distance.

The magnetic field downstream of the shock is therefore significantly different in those two regions. 
This emphasizes the fact that  the magnetic field derived from a one-zone modeling of the SED of the whole SNR
 (giving an average value of the magnetic field) is not directly comparable to the magnetic field derived from a filament in a specific region of the SNR.

\section{Conclusions}

The search for the X-ray counterpart of PSR\,J0855$-$4644 using \textit{XMM-Newton} led to the following conclusions:

\begin{enumerate}

\item  A point-like X-ray counterpart has been discovered in spatial coincidence with the position of the radio pulsar and a surrounding extended emission of non-thermal origin was found.
 The nebula has a maximum extent of the order of 150'' and the azimuthal profile has revealed two knots of emission (with and angular separation of $\sim$180$\degr$),
  suggesting possible jets around the pulsar. High spatial resolution observations with \textit{Chandra} might reveal the fine structure of the nebula.

\item A faint rim of non-thermal emission representing the South-Eastern part of the RX J0852.0$-$4622 SNR was detected. 
The filament in this region is  thicker than the bright and  thin X-ray filaments observed in the North-West of the SNR. 
Assuming that the thickness of the rim is dominated by the synchrotron losses, the downstream magnetic field is
24 $\mu$G and 63 $\mu$G respectively for the South-East and for the North-West rim. Those significantly different values
might indicate different acceleration and/or environmental conditions at the shock around the SNR.

\item By comparing  the X-ray column density on the pulsar and 
on different regions of the SNR with the  gas map (using $^{12}$CO and HI emission as  tracers), we 
concluded that both the RX J0852.0$-$4622 SNR and PSR\,J0855$-$4644  are in the foreground or at the interface of  the Vela Molecular Ridge resulting in an upper limit of d$\leq$ 900 pc. 
Although the pulsar and the SNR may be at similar distances, they are not likely to be associated. 

\item The new estimate of the pulsar distance completes our knowledge of the nearby pulsar population.
If the distance of the pulsar is confirmed through e.g. radio parallax measurements, PSR\,J0855$-$4644 would be 
the second most energetic pulsar (after the Vela pulsar) within 1 kpc and may contribute to the observed e$^{-}$/e$^{+}$ spectrum.

\end{enumerate}

 \begin{center}
\begin{small}ACKNOWLEDGMENTS          
\end{small}         
\end{center}

F.A. would like to thank Isabelle Grenier for discussions on the nearby interstellar medium and David Smith for useful discussion on the NE2001 Galactic electron distribution model. 
The help from Konrad Dennerl is also greatly appreciated to solve the calibration issues with the \textit{XMM-Newton} PN camera.

 \begin{appendix}
 \section{Calibration issues with the PN camera}
\label{sect:appendix}
\begin{figure}[t]

\includegraphics[bb= -10 0 455 905,clip,angle=-90,width=8.5cm]{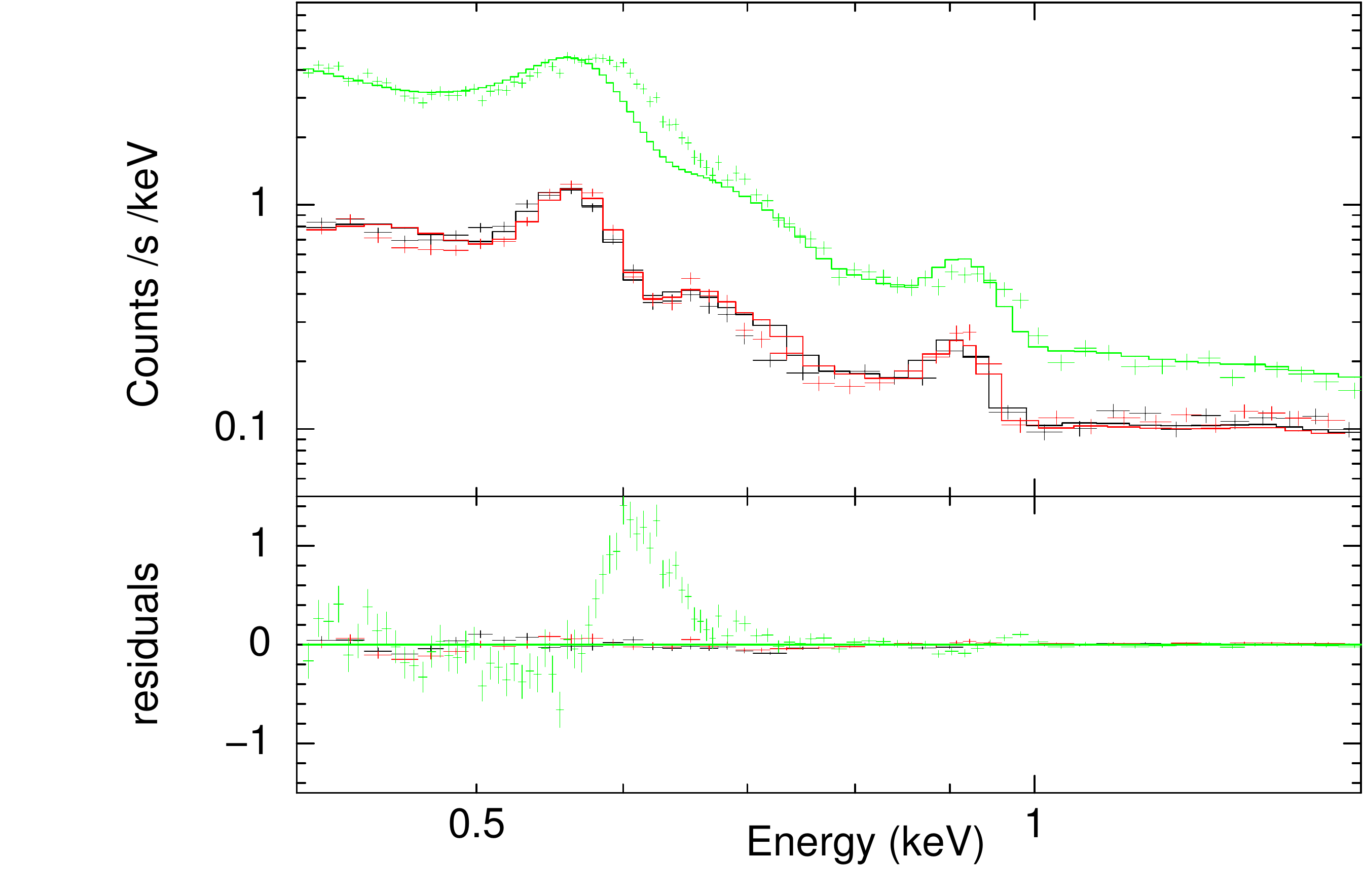}  \vspace{0mm} \\
\includegraphics[bb= -10 0 510 905,clip,angle=-90,width=8.5cm]{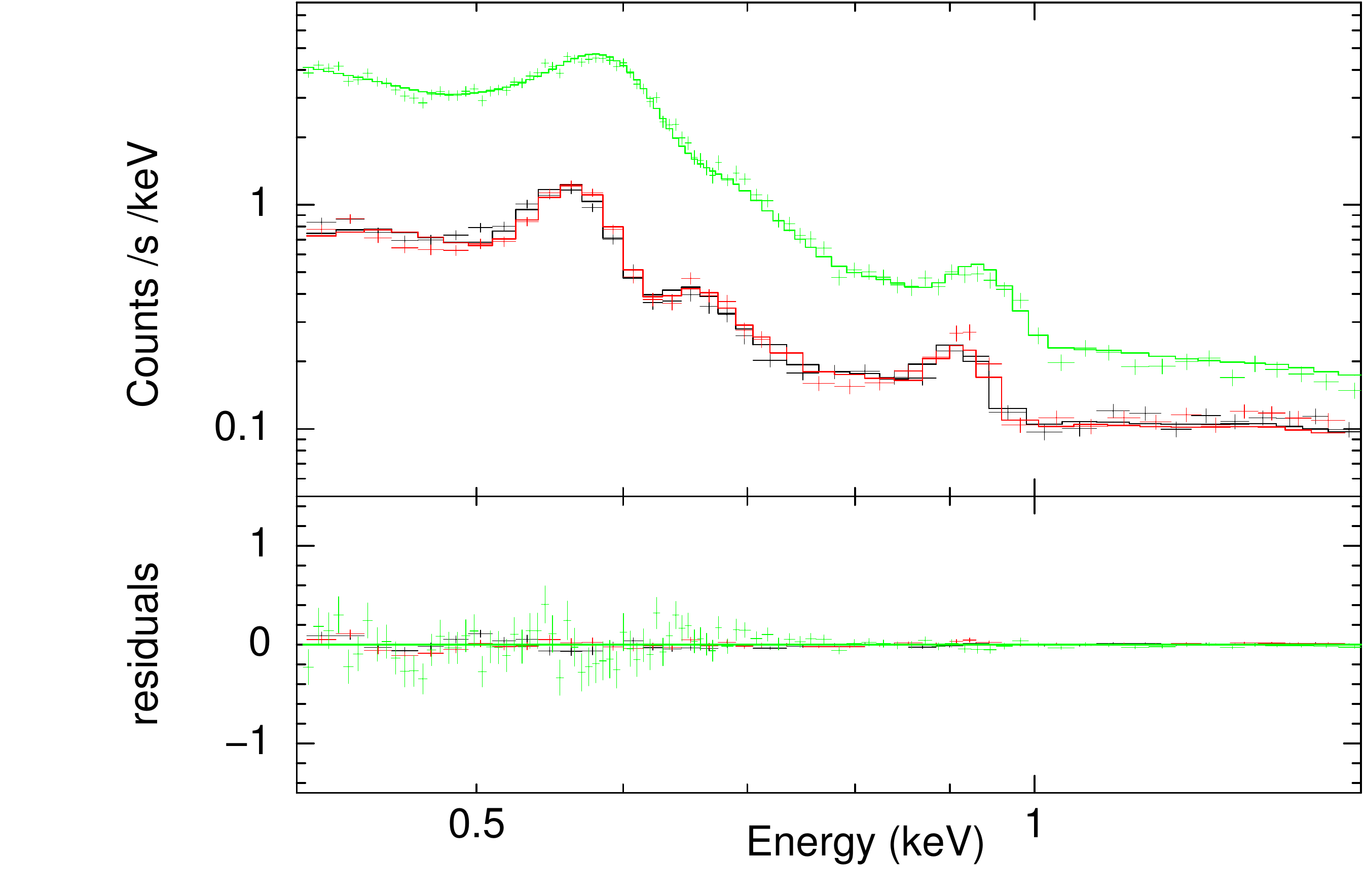}

\vspace{-0.3cm}
\caption{
\footnotesize
\textit{Top}: X-ray spectrum of the nebula extracted from a 15-150'' annulus region shown in Fig. \ref{profiles} focused on the thermal emission
 with MOS1, MOS2 and PN spectra represented in black, red and green respectively. The same spectral model as in Fig. \ref{spec} is overlaid.
\textit{Bottom}:   Same spectra as in the Top panel but with the PN response matrix modified by the \textit{gain} response model implemented in Xspec.
 }
\label{PNspec}
\end{figure}

The thermal emission from the Vela SNR is extremely bright at low energies (E$<$1 keV), an energy range where the PN camera is more sensitive than the MOS. 
With such a statistics we are limited by our knowledge of the calibration of the camera. 
Here we have observed  a deficiency in the low energy calibration of the spectral response of the PN camera. This effect is most visible around 0.6 keV in Fig. \ref{PNspec}
 where the Oxygen lines can not be reproduced by the APEC thermal model. 

This is a known calibration 
issue\footnote{More information on this calibration issue can be found here: 
http://www2.le.ac.uk/departments/physics/research/src/Missions/xmm-newton/technical/cal-meetings$\sharp$madrid-0310} 
that is being investigated by the \textit{XMM-Newton} PN calibration team. There are no known similar calibration issues for the MOS instruments.
Following their recommendation, we allowed for an empirical shift in energy in the PN response file using the \textit{gain} response model implemented in Xspec.
The fitted energy shift is of the order of 15 eV. Similar best-fit parameters where obtained when fitting independently the MOS and response corrected PN spectra. 
This confirms that the correction of the response matrix does not alter the physical parameters that we derive from the spectrum.   
 In the spectral analysis of the article we have therefore used the MOS1, MOS2 and response corrected PN spectra.
  
\end{appendix}

\end{document}